\documentclass[twocolumn,trackchanges,appendixfloats,captions=figureheading]{aastex61}
	
\usepackage[T1]{fontenc}
\usepackage[utf8]{inputenc}
\usepackage{amsmath}
\usepackage{multirow}
\usepackage{booktabs}
\usepackage{float}
\usepackage{xspace}
\usepackage[caption=false]{subfig}
\usepackage{placeins}

\newcommand{\kms}{km\,s$^{-1}$\xspace}
\newcommand{\jybeam}{Jy\,beam$^{-1}$\xspace}
\newcommand{\Kkmspc}{K\,km\,s$^{-1}$\,pc$^2$\xspace}
\newcommand{\sqcm}{cm$^{-2}$\xspace}
\newcommand{\ergs}{erg\,s$^{-1}$\xspace}
\newcommand{\Msun}{M$_\odot$\xspace}
\newcommand{\Msunyr}{M$_\odot$\,yr$^{-1}$\xspace}
\newcommand{\Msunkms}{M$_\odot$\,km\,s$^{-1}$\xspace}
\newcommand{\co}[2]{CO(#1--#2)\xspace}
\newcommand{\hcn}{HCN~(4-3)\xspace}
\newcommand{\hco}{HCO$^+$~(4-3)\xspace}
\newcommand{\cs}{CS~(7-6)\xspace}
\newcommand{\casa}{\textsc{casa}\xspace}

\begin{document}

\received{-}
\revised{-}
\accepted{26 Jun 2019}
\published{-}
\submitjournal{\apj}

\title{The Molecular Outflow in NGC\,253 at a Resolution of Two Parsecs}

\author[0000-0003-1104-2014]{Nico Krieger}
	\altaffiliation{email: krieger@mpia.de}
	\affiliation{Max-Planck-Institut f\"ur Astronomie, K\"onigstuhl 17, 69120 Heidelberg, Germany}
\author{Alberto D. Bolatto}
    \affiliation{Department of Astronomy, University of Maryland, College Park, MD 20742, USA}
\author{Fabian Walter}
    \affiliation{Max-Planck-Institut f\"ur Astronomie, K\"onigstuhl 17, 69120 Heidelberg, Germany}
    \affiliation{National Radio Astronomy Observatory, P.O. Box O, 1003 Lopezville Road, Socorro, NM 87801, USA}
\author{Adam K. Leroy}
    \affiliation{Department of Astronomy, The Ohio State University, 4055 McPherson Laboratory, 140 West 18th Ave, Columbus, OH 43210, USA}
\author{Laura K. Zschaechner}
    \affiliation{Finnish Center for Astronomy with ESO, FI-20014 Turun yliopisto, Finland}
    \affiliation{University of Helsinki, P.O. Box 64, Gustaf H\"allstr\"omin katu 2a, FI-00014 Helsingin yliopisto, Finland}
\author{David S. Meier}
    \affiliation{New Mexico Institute of Mining and Technology, 801 Leroy Place, Socorro, NM 87801, USA}
    \affiliation{National Radio Astronomy Observatory, P.O. Box O, 1003 Lopezville Road, Socorro, NM 87801, USA}
\author{J\"urgen Ott}
    \affiliation{National Radio Astronomy Observatory, P.O. Box O, 1003 Lopezville Road, Socorro, NM 87801, USA}
\author{Axel Wei\ss}
    \affiliation{Max-Planck-Institut f\"ur Radioastronomie, Auf dem H\"ugel 69, 53121 Bonn, Germany}
\author{Elisabeth A.C. Mills}
    \affiliation{Physics Department, Brandeis University, 415 South Street, Waltham, MA 02453}
\author{Rebecca C. Levy}
    \affiliation{Department of Astronomy, University of Maryland, College Park, MD 20742, USA}
\author{Sylvain Veilleux}
    \affiliation{Department of Astronomy, University of Maryland, College Park, MD 20742, USA}
\author{Mark Gorski}
    \affiliation{Department of Physics and Astronomy, University of Western Ontario, London, Ontario N6A 3K7, Canada}

\shorttitle{The Molecular Outflow in NGC\,253 at a Resolution of Two Parsecs}
\shortauthors{Krieger et al.}

\begin{abstract}
We present 0.15\arcsec ($\sim2.5$\,pc) resolution ALMA \co32 observations of the starbursting center in NGC\,253. Together with archival ALMA \co10 and \co21 data we decompose the emission into a disk and non--disk component. 
We find $\sim7-16\%$ of the CO luminosity to be associated with the non-disk component ($1.2-4.2 \times 10^7$\,\Kkmspc). The total molecular gas mass in the center of NGC\,253 is $\sim3.6\times10^8$\,M$_\odot$ with $\sim0.5\times10^8$\,M$_\odot$ ($\sim15\%$) in the non-disk component. These measurements are consistent across independent mass estimates through three CO transitions. 
The high-resolution \co32 observations allow us to identify the molecular outflow within the non-disk gas. Using a starburst conversion factor, we estimate the deprojected molecular mass outflow rate, kinetic energy and momentum in the starburst of NGC\,253. The deprojected molecular mass outflow rate is in the range $\sim 14-39$\,\Msunyr with an uncertainty of 0.4\,dex. The large spread arises due to different interpretations of the kinematics of the observed gas while the errors are due to unknown geometry. The majority of this outflow rate is contributed by distinct outflows perpendicular to the disk, with a significant contribution by diffuse molecular gas. This results in a mass loading factor $\eta = \dot{M}_\mathrm{out} / \dot{M}_\mathrm{SFR}$ in the range $\eta\sim8-20$ for gas ejected out to $\sim300$\,pc. 
We find the kinetic energy of the outflow to be $\sim2.5-4.5\times10^{54}$\,erg and $\sim0.8$\,dex typical error which is $\sim0.1$\% of the total or $\sim8$\% of the kinetic energy supplied by the starburst. The outflow momentum is $4.8-8.7\times10^8$\,\Msunkms ($\sim0.5$\,dex error) or $\sim2.5-4$\% of the kinetic momentum released into the ISM by feedback. 
The unknown outflow geometry and launching sites are the primary source of uncertainty in this study.
\end{abstract}



\section{Introduction}

Outflows driven by star formation are thought to be a crucial driver of galaxy evolution. Strong stellar feedback caused by high star formation rate densities can launch outflows of ionized, neutral and molecular gas that potentially can escape the main body of a galaxy. Consequently, such outflowing gas removes the potential fuel for future star formation. Therefore, outflows can suppress and quench star formation, as also demonstrated by theoretical predictions and simulations \citep[e.g.][]{1986ApJ...303...39D,2017MNRAS.466.1213K,2018ApJ...857..116M}. Depending on the velocity of the outflow and a galaxy's escape velocity, outflowing gas can be re--accreted at later cosmic times (the so--called `galactic fountain') or leave the system altogether. This process thus has the potential to enrich the galactic disk and circum--galactic medium with heavy metals \citep[e.g.][]{Oppenheimer:2006eq,2010MNRAS.406.2325O,Hopkins:2012ez,Christensen:2018ka}.

Galactic outflows are a multi--phase phenomenon and are observed across the electro--magnetic spectrum from X-ray \citep[e.g.][]{2007ApJ...658..258S}, UV \citep[e.g.][]{2005ApJ...619L..99H}, optical like H$\alpha$ \citep[e.g.][]{2009ApJ...696..192W} to IR \citep[e.g.][]{2009ApJ...700L.149V}, cold dust \citep[e.g.][]{2010A&A...518L..66R}, PAH emission \citep[e.g.][]{2006ApJ...642L.127E}, and sub-millimeter to radio including H\textsc{i} \citep[e.g.][]{2013Natur.499..450B,2015ApJ...814...83L,Lucero:2015if}. Typically, large-scale outflow features at high relative velocity (100s-1000s \kms) are observed in the ionized and neutral gas, whereas molecular outflows often appear as smaller, more compact features \citep{Strickland:2002kp,Westmoquette:2011bp}. The latter are nonetheless important as they dominate the mass budget \citep{2015ApJ...814...83L}. In some galaxies, the gas phases seem to be stratified with an inner ionized outflow cone, a surrounding neutral shell, and molecular gas situated along the outer edge \citep[e.g.][]{2015ApJ...801...63M}. Typically, the outflows originate from an extended region, so the apparent outflow cone has its tip cut-off. 

Molecular outflows are thus closely intertwined with feedback processes and star formation. The high-resolution structure and kinematic properties of (molecular) outflows are not studied in great detail yet, primarily due to the lack of high resolution and sensitivity observations. Starburst galaxies are the obvious target to study star formation-driven outflows due to the high star formation rates (SFR) in these system. Consequently, molecular outflows have been studied over the past years in a few nearby starbursts : 
M\,82 \citep{2002ApJ...580L..21W,2015ApJ...814...83L}, NGC\,253 \citep{2013Natur.499..450B,2017ApJ...835..265W,2018ApJ...867..111Z}, NGC\,1808 \citep{2018ApJ...856...97S}, and ESO320-G030 \citep{2016A&A...594A..81P}.

NGC\,253 is one of the nearest starburst systems at a distance of 3.5\,Mpc \citep{Rekola:2005ha}. It is considered one of the prototypical starburst galaxies with a star formation rate surface density of $\Sigma_{SFR} \sim 10^2$\,\Msun\,yr$^{-1}$\,kpc$^{-2}$ in the nuclear region and a molecular depletion time that is $\tau^{mol}_{dep} \sim 5-25$ times lower than what is found in local disks \citep{Leroy:2015ds}. A galactic wind emerges from the central $\sim 200$\,pc of NGC\,253 that has been characterized in H$\alpha$, X-ray, as well as neutral and molecular gas emission \citep[e.g.][]{Sharp:2010jl,Turner:1985iy,Sturm:2011jb,Strickland:2000wd,Strickland:2002kp, Westmoquette:2011bp,2000ApJS..129..493H,2013Natur.499..450B,2017ApJ...835..265W}.
Due to the close proximity, starburst and galactic winds can be studied in detail and individual structures can be resolved.

Studies of the molecular gas phase in NGC\,253 showed that its central starburst is fueled by gas accretion along the bar \citep{2004ApJ...611..835P}. The molecular ISM in the nuclear region is structured in several clumps that show high temperatures of $\sim 50$\,K \citep{2004ApJ...611..835P,Sakamoto:2011et,2019ApJ...871..170M}. From earlier low resolution observations \citep[$>20$\,pc, e.g.][]{2006ApJ...636..685S,Sakamoto:2011et} to recent observations at high resolution ($8\,\mathrm{pc}\times5$\,pc in \citealt{2017ApJ...849...81A} and 2\,pc in \citealt{2018ApJ...869..126L}) the number of molecular clumps associated the starburst increased from $\sim5$ to 14. These studies find the clumps to be massive ($4-10 \times 10^4$\,\Msun), compact ($<10$\,pc), chemically rich (up to $>19$ molecules detected in the 0.8\,mm band) and hot (up to 90\,K). Each clump likely hosts an embedded massive star cluster \citep{2018ApJ...869..126L}. Further structures in the molecular gas are shells and bubbles blown up by feedback from  the intense star formation process. \citet{2006ApJ...636..685S} found two 100\,pc diameter superbubbles. \citet{2013Natur.499..450B} report molecular streamers\footnote{The term {\em streamer} here denotes structures with a high aspect ratio that are typically oriented roughly perpendicular to the disk and often show a velocity gradient.} originating from these shells with a lower limit to the outflow rate of $3-9$\,M$_\odot$\,yr$^{-1}$, about three times the star formation rate. This estimate was revisited by \citet{2018ApJ...867..111Z}, based on observations that show that the CO emission associated with the most prominent streamer is optically thick, increasing it to $25-50$\,\Msunyr.

As suggested by these studies, the outflow rate in NGC\,253 is factors of a few to potentially $>10$ larger than the star formation rate. Hence, the impact of the outflows on the amount of material lost from the molecular gas reservoir, and thus the lifetime of the starburst, is significant. The availability of new data makes it interesting to revisit the determination of the mass outflow rate in NGC\,253, while also removing some limitations of  previous determinations. \citet{2013Natur.499..450B} estimated the outflow rate from a few massive molecular streamers, but did not include potential diffuse outflowing gas. Also, resolution plays an important role in the ability to disentangle outflows from material in the starbursting disk. New ALMA band~7 observations provide excellent spatial resolution and reasonable surface brightness sensitivity. This information enables increasingly accurate determination of the total mass outflow rate, and its impact on the starburst.

In this work, we present ALMA \co32 observations carried out in cycle 3 and 4 that target the molecular gas in the central $\sim 750$\,pc of NGC\,253. Together with ancillary band 3 and 6 data from our previous work \citep{2013Natur.499..450B,2015ApJ...801...63M,Leroy:2015ds,2018ApJ...867..111Z}, we have an inventory of three CO lines to study the molecular gas in the starbursting disk and a kinematically different component that includes the outflow. By decomposing the detected emission, we aim to measure the total molecular gas outflow rate in NGC\,253 and improve upon previous less systematic results.

Throughout this paper, we adopt a distance of 3.5\,Mpc to NGC\,253 \citep{Rekola:2005ha} at which 1\arcsec\ corresponds to 17\,pc. We also define the ``center'' of the nuclear region of NGC\,253 to be the kinematic center at $\alpha, \delta = 00^h47^m33.134^s, -25^\circ17^m19.68^s$ as identified in \citet{MullerSanchez:2010dr}.
The paper is structured as follows: In section~\ref{section: data reduction}, we describe observational setup and data reduction, and show the results in the form of channel maps, moment maps and position-velocity diagrams. Our approach on separating gas in the star-forming disk from potentially outflowing gas is laid out in section~\ref{section: disk separation}. Section~\ref{section: results separated disk/non-disk} discusses the derived quantities such as CO luminosities, molecular gas masses, outflow rate, kinetic energy and momentum. Our conclusions are summarized in section~\ref{section: summary}.


\section{Data Reduction and Imaging}\label{section: data reduction}


\floattable
\begin{deluxetable*}{lccc}
	\tablewidth{\linewidth}
	\tablecaption{Details of the datasets used in this analysis.
	\label{table: used datasets}}
	\tablehead{\colhead{} & \colhead{\co10} & \colhead{\co21} & \colhead{\co32}}
	\startdata
	ALMA ID				& 2011.1.00172.S		 	 & 2012.1.00108.S			  & 2015.1.00274.S\\
	spatial resolution	& $1.85\arcsec \times 1.32\arcsec$ 	 & $1.70\arcsec \times 1.02\arcsec$	  & $0.17\arcsec \times 0.13\arcsec$\\
	                    & 31.4\,pc $\times$ 22.4\,pc & 28.8\,pc $\times$ 17.3\,pc & 2.9\,pc $\times$ 2.2\,pc\\
	spectral resolution	& 5.0\,\kms					 & 5.0\,\kms				  & 2.5\,\kms\\
	RMS noise per channel & 1.99\,m\jybeam             & 2.19\,m\jybeam             & 0.81\,m\jybeam\\
	                    & 75\,mK					 & 29\,mK					  & 0.37\,K\\
	\enddata
\end{deluxetable*}

\subsection{Data reduction}

The data presented in this paper are based on observations in ALMA cycles~2, 3 and 4 in bands~3, 6 and 7 that cover the redshifted emission in NGC\,253 of \co10, \co21 and \co32 as well as other molecular lines.
For data reduction and imaging of the band~3 and 6 data see \citet{2013Natur.499..450B}, \citet{Leroy:2015ds}, \citet{2015ApJ...801...63M} and \citet{2018ApJ...867..111Z}. Table~\ref{table: used datasets} gives an overview of the datasets used in this analysis.

For the band~7 observations, we tuned the lower side band to $342.0-345.8$\,GHz and the upper side band to $353.9-357.7$\,GHz (total bandwidth 7.6\,GHz) with 976.6\,kHz channel width (corresponding to 0.8\,\kms). We targeted the central $\sim 750$\,pc of NGC\,253 in a linear four pointing mosaic with two configurations of the 12\,m array (12\,m compact and 12\,m extended, half power beam width $\sim 17\arcsec$) and a five pointing mosaic of the 7\,m array (ACA, half power beam width $\sim 30\arcsec$). Additional single dish observations with the total power array (TP) recovers emission on large spatial scales. The baseline ranges covered by this setup are 8.9-49.0\,m, $15.1 - 783.5$\,m and $15.1 - 1813.1$\,m for the ACA and the two 12\,m setups, respectively.

The observations were carried out primarily in the first half of 2016 (TP: 07-Dec-2015 to 02-Aug-2016; ACA: 07-Dec-2015 to 23-Nov-2016; 12\,m compact configuration: 16-Apr-2016, 23-Apr-2016, 17-Jun-2016, 27-Jun-2016; 12\,m extended configuration: 30-Aug-2016, 03-Sep-2016). The total on-source observation time is $48^h45^m$ split across $27^h:23^m$ (TP), $14^h57^m$ (ACA), $2^h37^m$ (12\,m compact) and $3^h59^m$ (12\,m extended). The calibrators were: J0006-0623 (bandpass); J0038-2459 (complex gain); the asteroid Pallas (absolute flux density); J0104-2416, J0106-2718 (both WVR). Visibilities of the 12\,m data are calibrated using the ALMA cycle~3 pipeline in \casa 4.6.0 and the delivered calibration script. The other datasets are calibrated in \casa 4.7.2 and the cycle~4 pipeline.

In order to image the spectral lines, we subtract the continuum in the $U,V$ plane using a first order polynomial fitted to the channels that do not contain strong spectral lines. We reliably detect $>25$ lines in the range 342.0-345.0\,GHz and 353.9-357.7\,GHz beside the four strong lines \co32, \hcn, \hco and \cs. Most of these lines are weak and only detected in small spatial regions so they do not affect the overall continuum fit and subtraction.

\subsection{Imaging}\label{subsection: imaging}

Combined imaging of the interferometric data is done with the \texttt{tclean} task in \casa 5.4.0 which includes crucial bug fixes for ALMA mosaics\footnote{For details see NAASC memo 117 by the North American ALMA Science Center (NAASC) at \url{http://library.nrao.edu/public/memos/naasc/NAASC_117.pdf}.}. We regrid the visibilities during deconvolution to a spectral resolution of 2.5\,\kms. Applying a Briggs weighting scheme with robust parameter 0.5 results in a synthesized beam of $0.17\arcsec \times 0.13\arcsec$ (pixel scale $0.05\arcsec$). The images are cleaned to a level of $2.5 \times$ the RMS noise in line-free channels of $2.5 \times 0.81$\,m\jybeam ($2.5 \times 0.37$\,K) using a clean mask derived from a low resolution image of the compact 12\,m array \co32 data only.

We correct the cleaned images for the mosaic sensitivity pattern (mosaic primary beam response pattern), combine them with the TP images using \texttt{feather} and finally convert the units to brightness temperature.

For the final images, we do not consider the ACA data as they introduce large scale noise fluctuation towards the edge of the mosaic, which we attribute to decreasing sensitivity of the 12\,m data relative to the ACA data. These fluctuations obscure the regions where outflows have been found previously. This work requires accurate integrated flux measurements and correct representation of the small scale structure which are defined by single dish observations (TP) and long baselines (extended 12\,m), respectively. By checking the images without ACA data against the images including ACA data, we can confirm that neither the overall flux scale, nor the small scale structure is significantly altered.

Data products for \co10 are shown in \citet{2013Natur.499..450B}, \citet{2015ApJ...801...63M}, \citet{Leroy:2015ds} and \citet{2018ApJ...867..111Z} presents the \co21 data. Imaging results for \co32 are presented in the following section.

In order to keep the amount of detail and contrast in the high resolution data, we do not match the spatial resolution to that of the data with the lowest resolution, but perform our analysis at the native resolution of each dataset. All further steps work on the data cubes masked at $5.0 \sigma$ (cf. table~\ref{table: used datasets}) and further masks where necessary. For generating the masks, we do not consider the non-uniform noise level caused by the mosaic sensitivity pattern but use the per channel RMS noise in the center of the field of view.


\subsection{\co32 data presentation}\label{section: channel maps}

\begin{figure*}
	\centering
	\includegraphics[width=\linewidth]{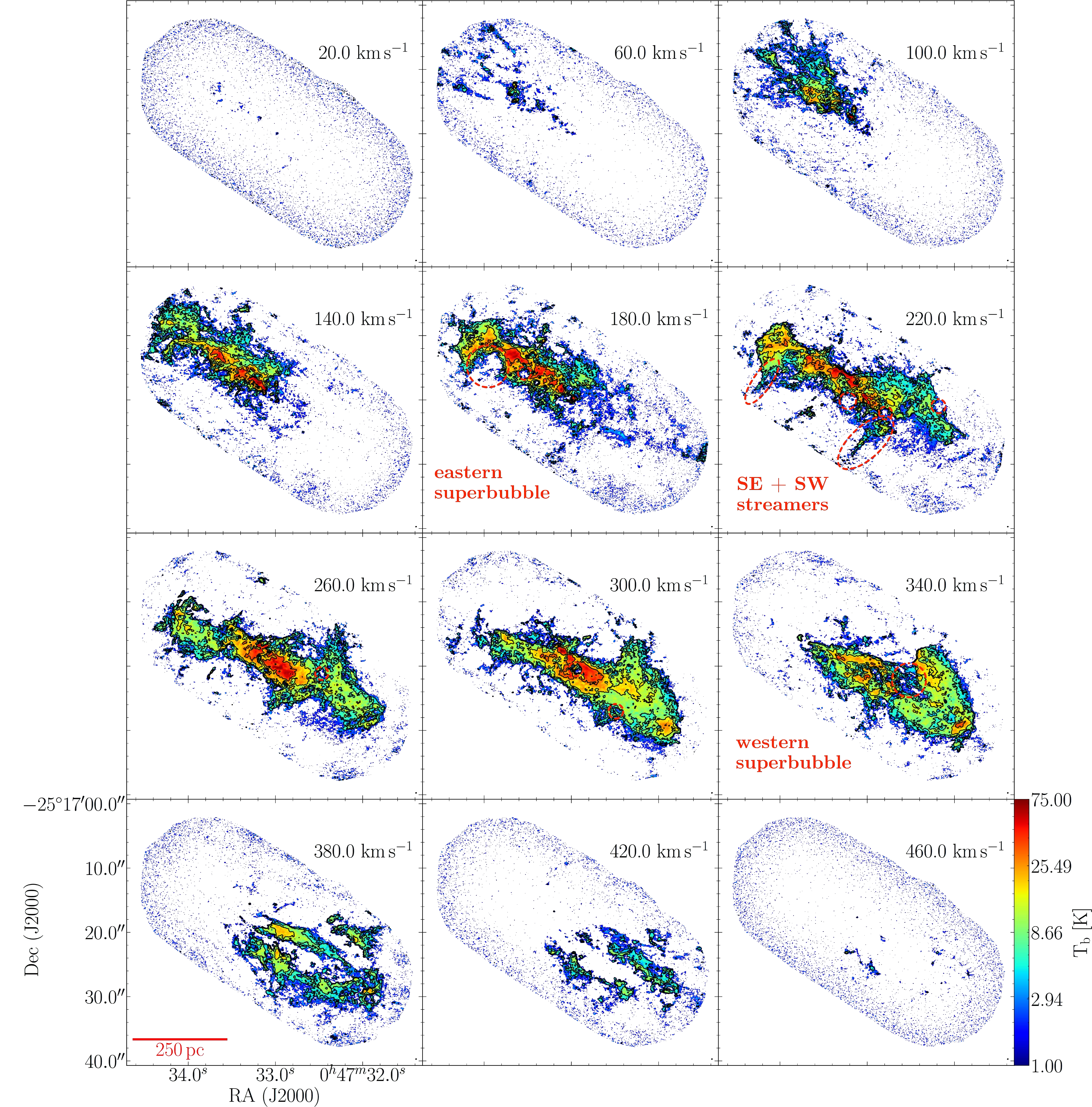}
	\caption{Channel maps of \co32 in NGC\,253. Every $16^{th}$ channel of 2.5\,\kms width is shown with the corresponding line-of-sight velocity ($\mathrm{v}_{sys} = 250$\,\kms) given in the upper right corner of each panel. The synthesized beam of $0.17" \times 0.13"$ is plotted in the lower right corner; it is hardly noticeable due to its small size. Contours are plotted at $10\sigma$, $20\sigma$, $40\sigma$, $80\sigma$ with an RMS noise of $\sigma = 0.37$\,K. Large structures are marked by dashed contours in those panels that show them most clearly. Further new shells are indicated by dashed circles.}
	\label{figure: CO channel map}
\end{figure*}

\begin{figure}
	\centering
	\includegraphics[width=\linewidth]{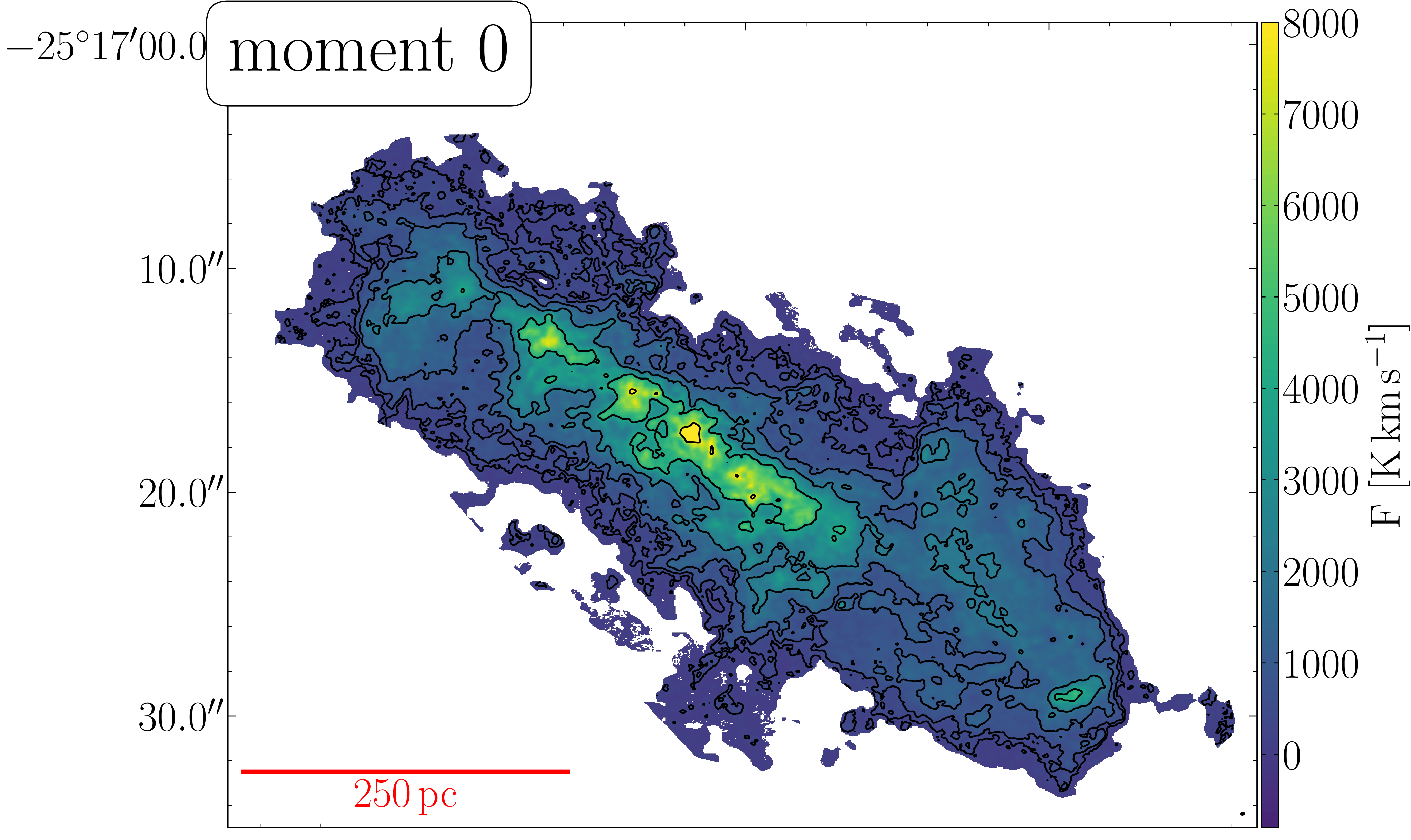}
	\includegraphics[width=\linewidth]{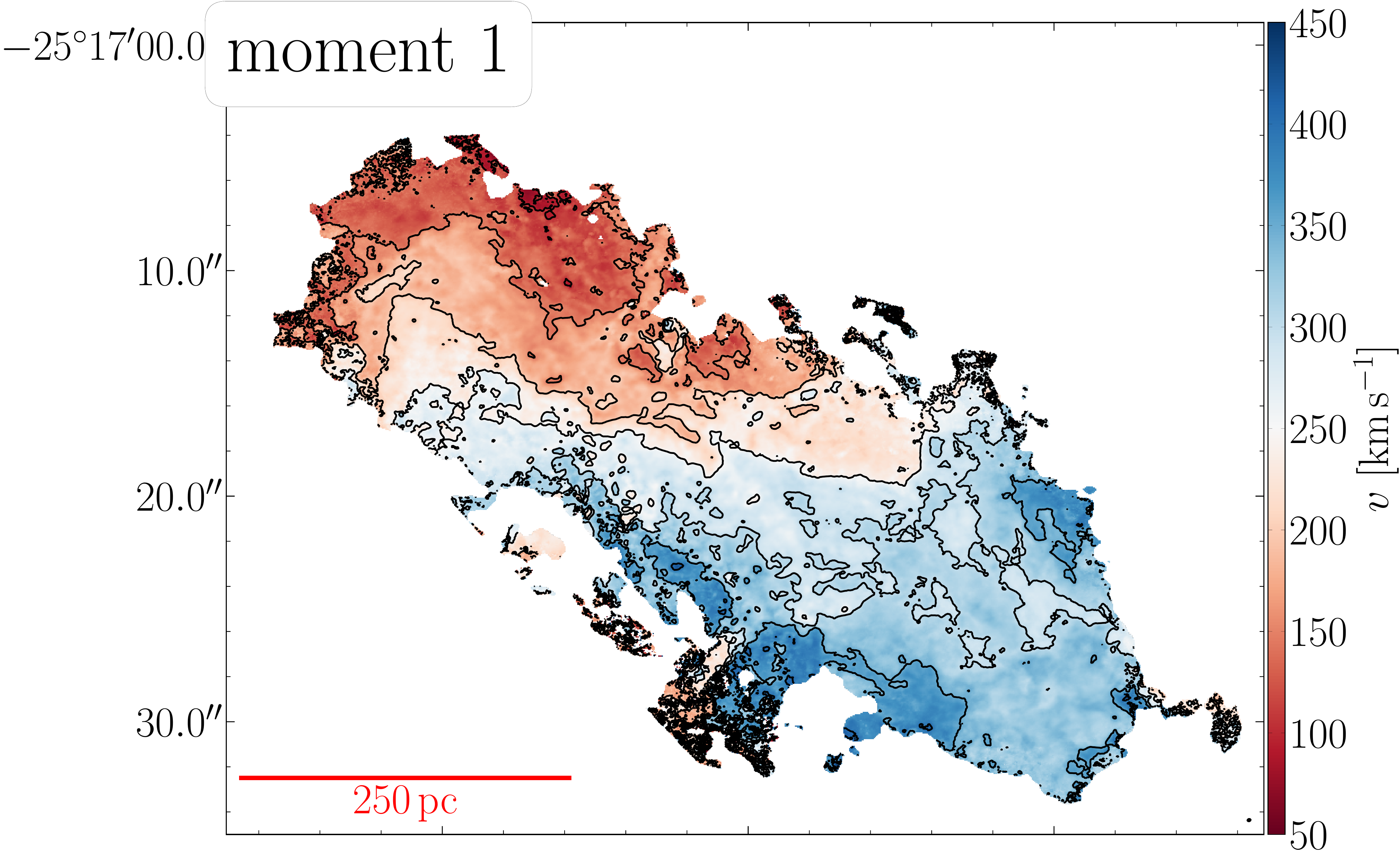}
	\includegraphics[width=\linewidth]{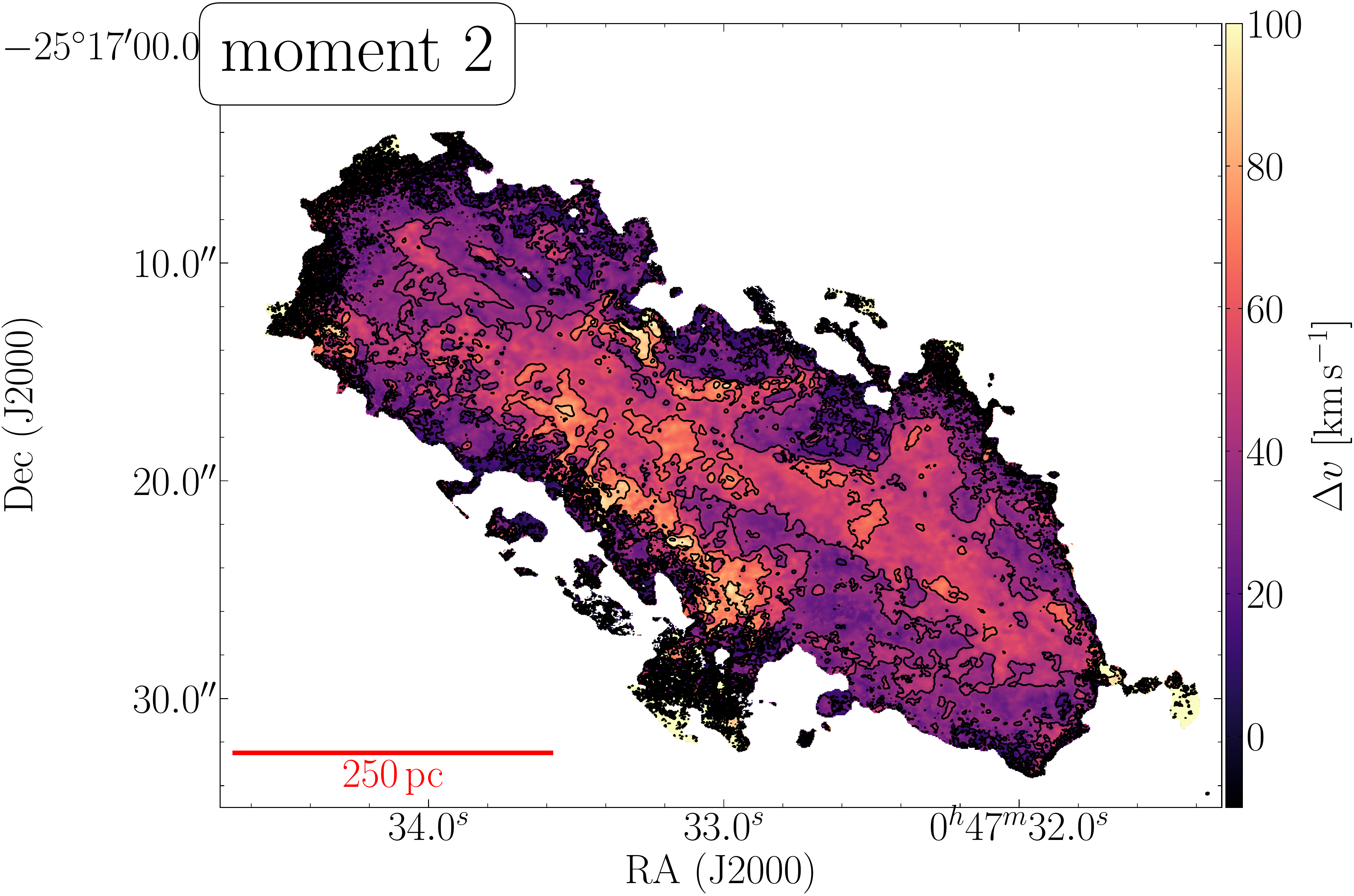}
	\caption{\co32 moment maps of NGC\,253. \emph{top}: Integrated intensity map (moment 0); contours are show from $250 - 8000$\,K\,\kms in factors of two. \emph{middle}: Velocity field (moment 1); contours are shown from 100\,\kms -- 400\,\kms in steps of 50\,\kms. \emph{bottom}: Moment 2 (corresponding to the velocity dispersion if the line profile would be Gaussian); contours are shown from 0\,\kms -- 100\,\kms in steps of 20\,\kms. The color scale is chosen to saturate a few regions with dispersions $>100$\,\kms. All maps are generated from the data cube masked a $5\sigma$ threshold per channel and confined to the collapsed clean mask to include only emission that has been processed by the clean algorithm.}
	\label{figure: CO moment maps}
\end{figure}

\begin{figure}
	\centering
	\includegraphics[width=\linewidth]{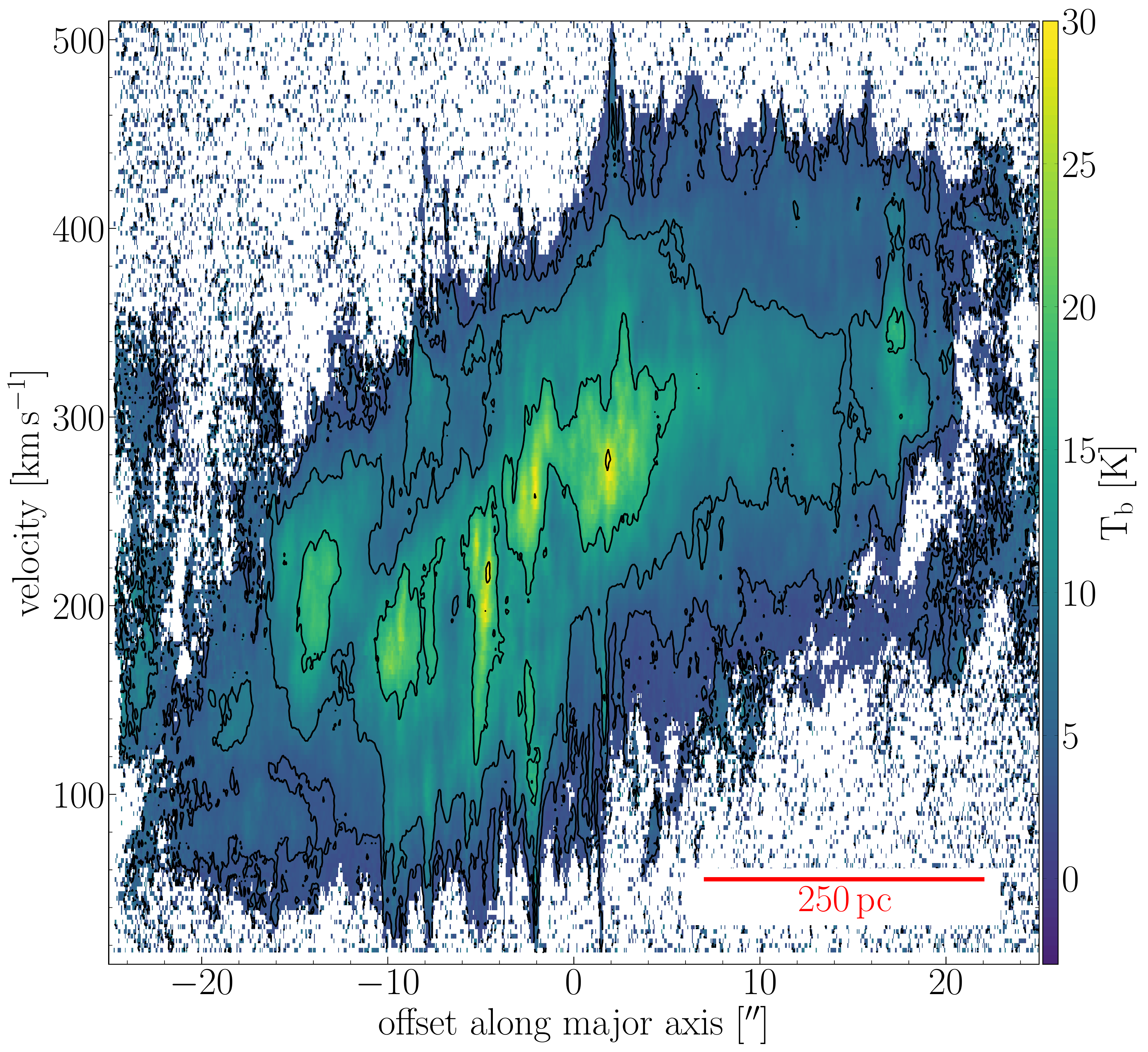}
	\caption{\co32 position-velocity diagram of NGC\,253 along the major axis centered on the kinematic center averaged over the full width of the field of view ($\sim 30\arcsec$). Pixel below $3\sigma$ are masked and contours are drawn at $10\sigma$, $20\sigma$, $40\sigma$, $80\sigma$ with an RMS noise of $\sigma = 0.37$\,K. Note the vertical spikes indicating high velocity dispersion due to outflowing gas.}
	\label{figure: co pV}
\end{figure}

In this section, we present the \co32 data in different representations. Channel maps (figure~\ref{figure: CO channel map}), moment maps (figure~\ref{figure: CO moment maps}) and a position-velocity (pV) diagram (figure~\ref{figure: co pV}) show the spatial and kinematic structures to be discussed and highlight the data quality.

Figure~\ref{figure: CO channel map} shows channel maps of the image cube. To retain the intrinsic resolution, only every 16$^{th}$ channel (40\,\kms spacing) is shown here. Besides the rotating disk of molecular gas, we clearly detect the prominent south-west (SW) streamer \citep{2017ApJ...835..265W} in the range $180 - 250$\,\kms (Fig.~\ref{figure: CO channel map}, panels 220\,\kms and 260\,\kms). Additional gas streamers are apparent between $\sim 60$\,\kms and $\sim 350$\,\kms towards north and south of the disk as can be seen for example in the panels at 260\,\kms or 340\,\kms. Several notable molecular shells are present between 180\,\kms and 340\,\kms. Beside the (super-)shells at the eastern (left) and western (right) edge of the map that have been previously identified by \citet{2006ApJ...636..685S} and \citet{2013Natur.499..450B}, further smaller shell--like structures are located along the molecular disk.

We calculate image moments (figure~\ref{figure: CO moment maps}) with \texttt{immoments} in \textsc{casa} for emission above $5\sigma$ for the moment 0 (integrated intensity) map, the moment~1 (intensity-weighted line-of-sight velocity) and moment~2 (intensity-weighted velocity dispersion) maps. Note that due to the complex line shapes, the moment~2 map does not directly correspond to velocity dispersion which is only the case for Gaussian line profiles. The maps are further constrained to the region defined by the collapsed clean mask to limit them to emission that has been processed by the clean algorithm.

Figure~\ref{figure: co pV} shows the kinematic structure of NGC\,253 as a pV diagram along the major axis of the disk ($\mathrm{PA} = 55^\circ$) averaged over the full width of the field of view ($\sim 30\arcsec$) centered on the kinematic center. The pV cut shows several high velocity dispersion structures extending from a rotating disk, indicative of outflows.


\section{Separating disk and non-disk emission}\label{section: disk separation}


\subsection{Separating disk and non-disk emission in position-position-velocity space}\label{subsection: ppV separation}

Our goal is to account for all the molecular wind, separating outflowing molecular gas from foreground or background disk emission. A clean separation in 2D position-position space cannot be easily accomplished due to the inclination of $78^\circ$ of NGC\,253. At this high inclination, outflows and disk emission are co-spatial in projection. Kinematic information from line-of-sight velocities, however, makes it possible to disentangle the outflow. Note that this becomes increasingly difficult as the velocity vector aligns with the plane of the sky, resulting in line-of-sight velocities that are systemic. From H$\alpha$ kinematic modeling the NGC\,253 outflow is approximately bi-conical with an axis normal to the disk and an opening angle of $\sim60^\circ$ \citep{Westmoquette:2011bp}, and thus the range of possible projection angles is large (see \citealt{2015ApJ...801...63M} for a sketch). Note that because the cone opening angle is larger than the angle between the axis of the cone and the plane of the sky, gas in the approaching and receding cones can have both blue- and red-shifted velocities with respect to systemic.

The launching of molecular gas occurs within the disk through star formation feedback, thus the outflows originate from the same location in position-position-velocity (ppV) space as disk molecular clouds. Outflows will therefore blend into the disk near their launching sites, which makes disentanglement increasingly difficult closer to the starburst region.

The complexity of systematically separating emission corresponding to the disk and the outflow in ppV space is challenging. Algorithmically, this separation is simpler in a lower dimensional space, obtained by slicing the data cube into a collection of 2D position-velocity diagrams. In what follows we identify kinematic components in these diagrams, which then we project back to 3D ppV space. In order to avoid introducing biases we model the large-scale disk velocity field and use this model as the basis of the kinematic separation.


\subsection{Definition of components}\label{subsection: definition disk non-disk}

Images of the center of NGC\,253 on large scales show an elongated gas structure (figure~\ref{figure: CO moment maps} top) with a regular velocity field (figure~\ref{figure: CO moment maps} middle) that roughly matches a rotating disk disturbed by streaming motions from a bar \citep[e.g.][]{2004ApJ...611..835P}. The elongated gas structure is consistent with a highly inclined disk of molecular gas, or possibly a ring-like structure as observed in other galaxies \citep[for example, NGC\,1512, NGC\,1808;][]{2016ApJ...823...68S,2018ApJ...857..116M}. 
Similar structures break up at higher spatial resolution into two embracing spiral arms or complex non-closed orbits in the Milky Way center \citep{2015MNRAS.453..739K,2016MNRAS.457.2675H,2018MNRAS.481....2S}. 

Superimposed on this large scale structure, there are smaller features that are not part of the large-scale pattern of rotation and streaming motions. Some of them have high aspect ratios in channel maps and line-of-sight velocity gradients, both typical for outflows. Local deviations from the large-scale velocity field can also be due to infalling gas, or clumps of gas that do not follow the global pattern due perhaps to a cloud-cloud collision.

Henceforth, we will refer to the bulk of the molecular gas that moves according to the large-scale velocity field in the central regions of NGC\,253 as the \emph{disk}. We assume this large-scale velocity field consists of rotation and streaming motions. The term \emph{non-disk} refers to any gas that is not following the ppV structure of the disk. By this definition, non-disk gas encompasses material from features that may be attributed to a variety of physical processes, including outflow and infall. 

Structures of outflowing gas are frequently referred to by names that describe their kinematic or spatial appearance, such as ``streamer''. We will use the term outflow to denote localized structures with morphology and kinematics consistent with gas moving away from the disk, as inferred from their location in ppV space. Typical signatures are a velocity that is inconsistent with rotation in the plane of the disk, and a high aspect ratio oriented roughly perpendicular to the disk major axis. Note that similar kinematic and structural properties can arise in infalling gas clouds. We will assume that all molecular gas with these characteristics is outflowing, which is likely the case for the majority of the material in NGC\,253.


\subsection{Position-velocity slicing}\label{subsection: slicing}
Kinematic analyses typically depend on high signal-to-noise ratios (SNR) because faint features can easily drown in noisy spectra. As a trade-of between necessary high SNR and also trying to include as much faint emission as possible, we conduct the following analysis on data cubes masked at the $5\sigma$ level (cf. table~\ref{table: used datasets}).

\begin{figure}
	\centering
	\includegraphics[width=\linewidth]{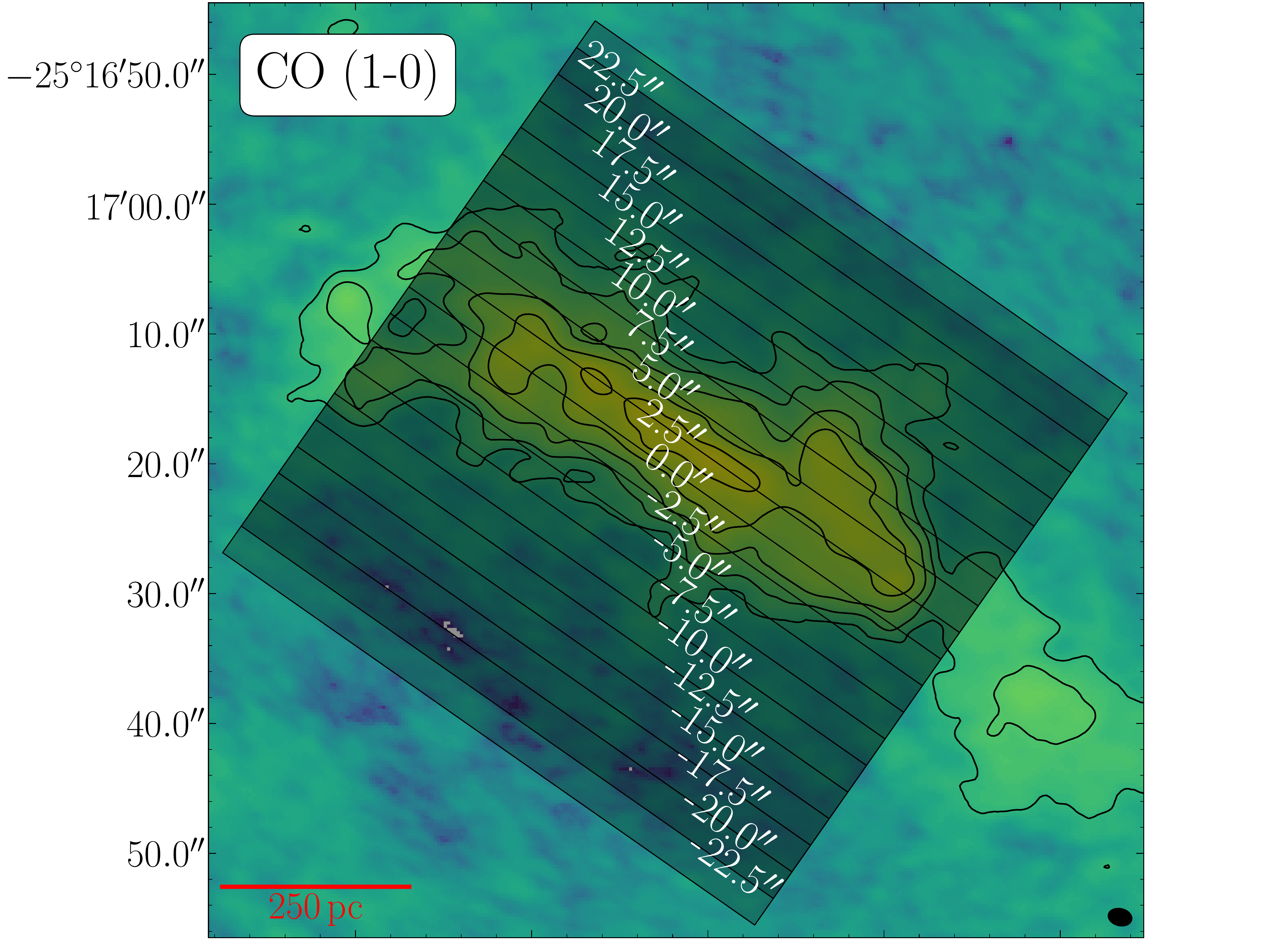}
	\includegraphics[width=\linewidth]{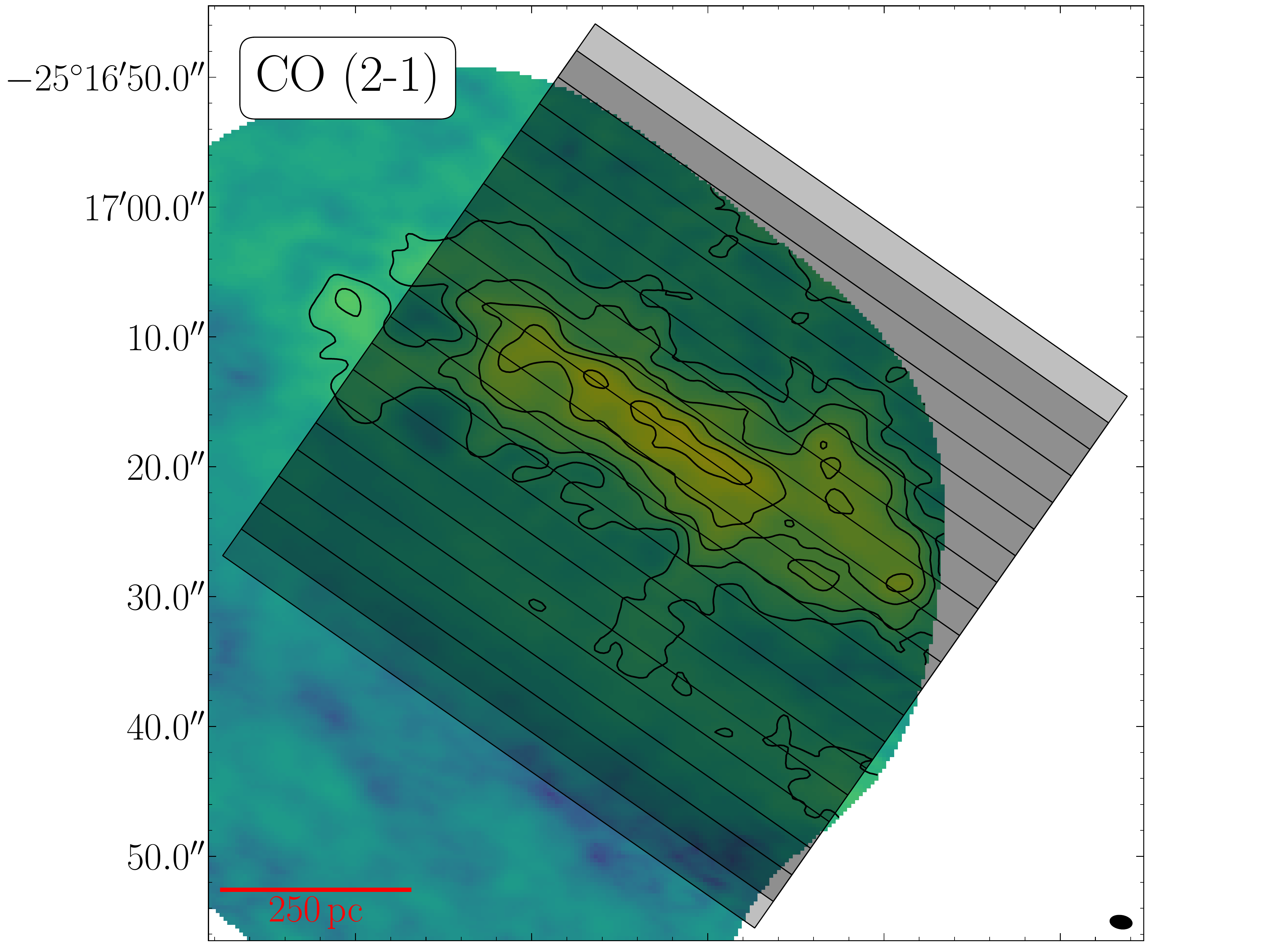}
	\includegraphics[width=\linewidth]{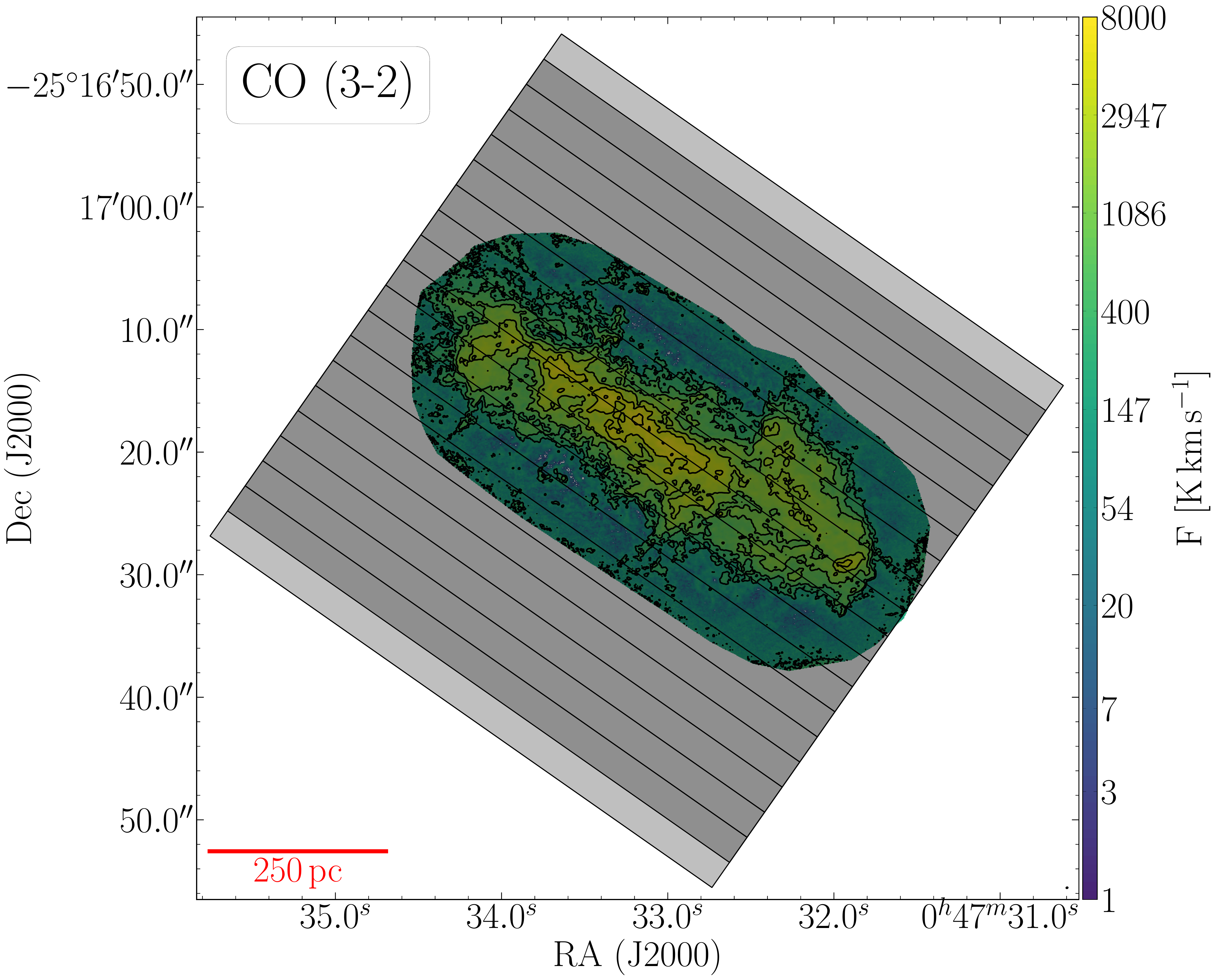}
	\caption{Size and orientation of the position-velocity slices overlaid on the integrated intensity image of \co10 (top), \co21 (middle) and \co32 (bottom). Each slice is $5.0\arcsec$ wide and overlaps adjacent slices by $2.5\arcsec$.}
	\label{figure: slice positions}
\end{figure}

We split the ppV cubes into position-velocity (pV) slices along the major axis of NGC\,253 as shown in figure~\ref{figure: slice positions}. The slices assume the kinematic center is $\alpha,\delta = 00^h47^m33.134^s, -25^\circ17^m19.68^s$ \citep{MullerSanchez:2010dr}, and are oriented along the major axis of the projected CO emission with $\mathrm{PA} = 55^\circ$. The area sliced is chosen to cover the region for which we have overlapping \co10, (2--1) and (3--2), and also cover the full length of the SW streamer outflow feature \citep[17.5\arcsec,][]{2017ApJ...835..265W}.

These requirements are fulfilled by slices of $50\arcsec$ (850\,pc) length (major axis) and covering $50\arcsec$ (850\,pc) along the minor axis (figure~\ref{figure: slice positions}). To reduce the problems introduced by splitting features across slices, each slice is $5.0\arcsec$ (85\,pc) wide, and we overlap slices by half their width ($2.5\arcsec$, 42\,pc). A sample pV diagram is shown in figure~\ref{figure: sample slice} for the central slice, which runs along the major axis (offset $0.0\arcsec$). A complete set of pV diagrams is given in appendix~\ref{appendix: all pVs}. The resolution differences between our three transitions, a factor of $\sim 100$ in beam solid angle, are apparent in figure~\ref{figure: sample slice}. In the high angular resolution \co32, small features with large linewidth are common. These features are blurred out in the lower resolution \co10 and (2--1).

\subsection{Modeling the disk}
We derive a model for the velocity of the disk component from the \co10 observations using the kinematic fitting tool \texttt{diskfit} \citep{2007ApJ...664..204S,2010MNRAS.404.1733S,2015arXiv150907120S}. Because the \co10 observations cover the largest area among our observations, we use them to derive the model; the additional information provided by the \co21 and/or \co32 data is negligible in terms of the bulk motions of the gas. We obtain a \co10 velocity field by computing the first moment of the cube after masking it at $20\sigma$ (1.26\,K), in order to represent the velocity of the bright emission.
We show the details of the fit parameters and a comparison to the \co10 velocity field in Appendix~\ref{appendix: model}.

In each pV slice, we use the velocity profile of the \texttt{diskfit} model to define the local disk velocity. We consider the CO emission consistent with the disk component of the emission when the velocity difference is within the local observed velocity range, $\Delta v$. This velocity range varies spatially and depends on distance $x$ from the major axis, increasing towards the center due to the combined effects of higher intrinsic velocity dispersion and projection.
For the success of this analysis, it is crucial that $\Delta v$ is broad enough to cover the observed velocity range of the disk but also narrow enough in order to not classify potential outflows as disk. The definition of $\Delta v$ is thus a crucial source of uncertainty for the derived quantities. We parametrize the velocity range of the disk as

\begin{equation}
	\Delta v\,(x) = 120\,\exp \left( - \left(\frac{x}{2.5}\right)^2 \right) + 100,
	\label{equation: delta v}
\end{equation}

\noindent with $\Delta v$ in km\,s$^{-1}$ and $x$ in arcsec. We find this empirical relation to fit the pV data best and small variations of order 10-20\% already show noticeable mismatch as is discussed in appendix~\ref{appendix: disk velocity range}. Note that the parameters (120, 2.5 and 100) in equation~\ref{equation: delta v} are visually selected to fit the fit pV diagrams as best as possible. The quality of this definition can be assessed from figure~\ref{figure: sample slice} and appendix~\ref{appendix: all pVs}: The velocity ranges are wide enough to include obvious emission of the disk but do not extend into the kinematically distinct features (potential outflows) that appear as spikes. This is most apparent for \co32 as this line offers the highest spatial resolution. The effect of a 10\% change in the velocity range $\Delta v$ correspond to up to 0.1\,dex variations in the derived quantities (cf. appendix~\ref{appendix: disk velocity range}).

\subsection{Selecting the components}
We use the modelled velocity field and the $\Delta v$ relation together to define a ``disk mask'' over ppV space, corresponding to emission that is consistent with disk rotation. We show in figure~\ref{figure: sample slice} the central pV slices for \co10, (2--1) and (3--2). We show in Appendix~\ref{appendix: all pVs} the complete set of pV slices.

Note that the CO emission extends beyond the disk mask. These extensions are not symmetric, and due to non-disk gas and projection effects. At $\sim78^\circ$ inclination gas flowing perpendicular to the galaxy disk towards the south (negative slice offsets) is primarily approaching us and seen at lower velocities relative to the disk emission. Similarly, outflow emission toward the north is primarily at velocities higher than the disk emission. Consequently, emission in pV slices shifts from lower to higher velocities relative to the disk model when the offset from the major axis increases (see figure~\ref{figure: all pV diagrams} in appendix~\ref{appendix: all pVs}). We designed the disk mask to be wide enough to capture the disk emission but exclude the asymmetric component caused by outflows.

We define a ``non-disk'' mask that is the mathematical complement of the disk mask, with the addition of removing emission from known sources (portions of the spiral arms) that were not included in our model of the central disk and are not of interest for this analysis.

\begin{figure}
	\centering
	\includegraphics[width=\linewidth]{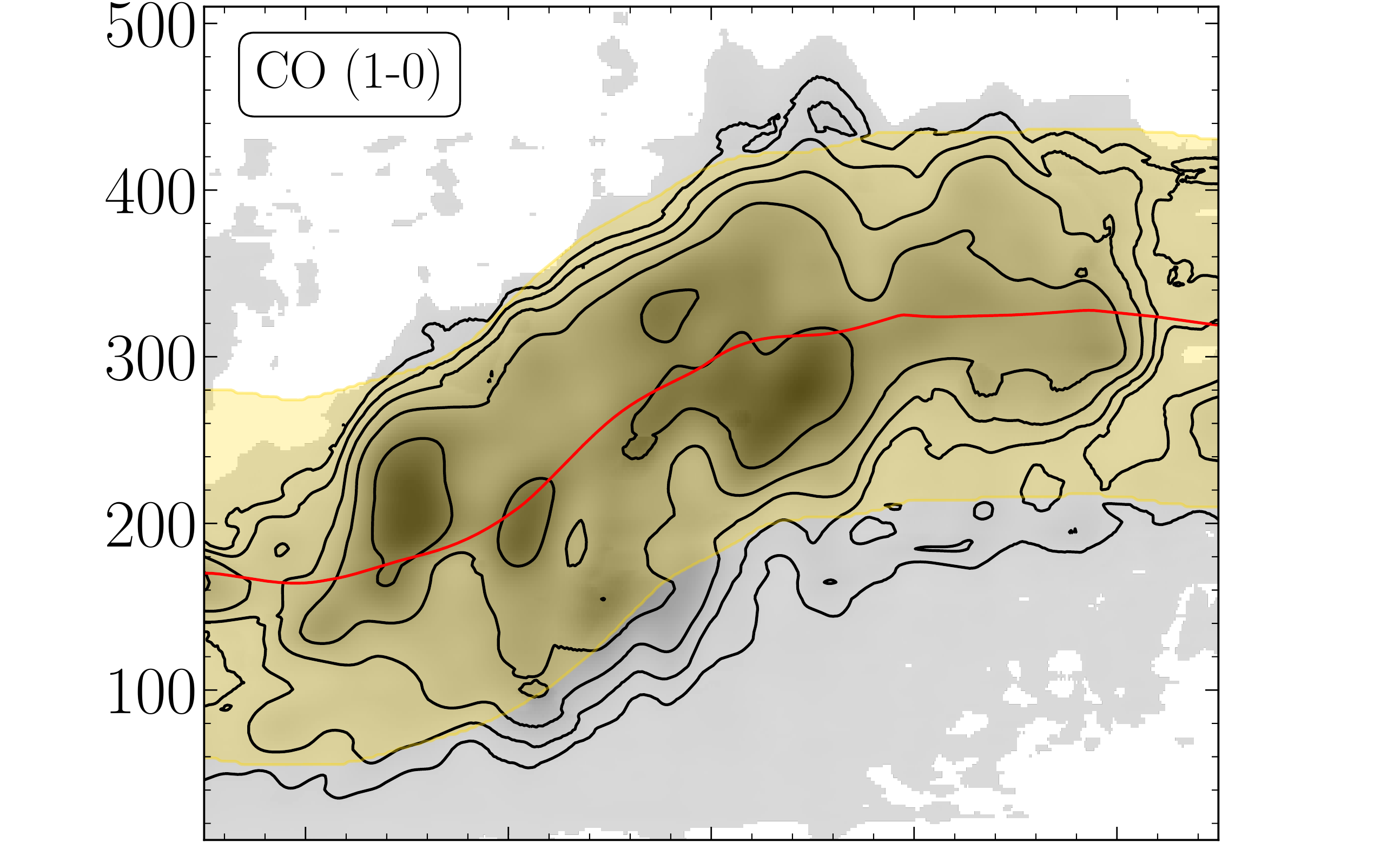}
	\includegraphics[width=\linewidth]{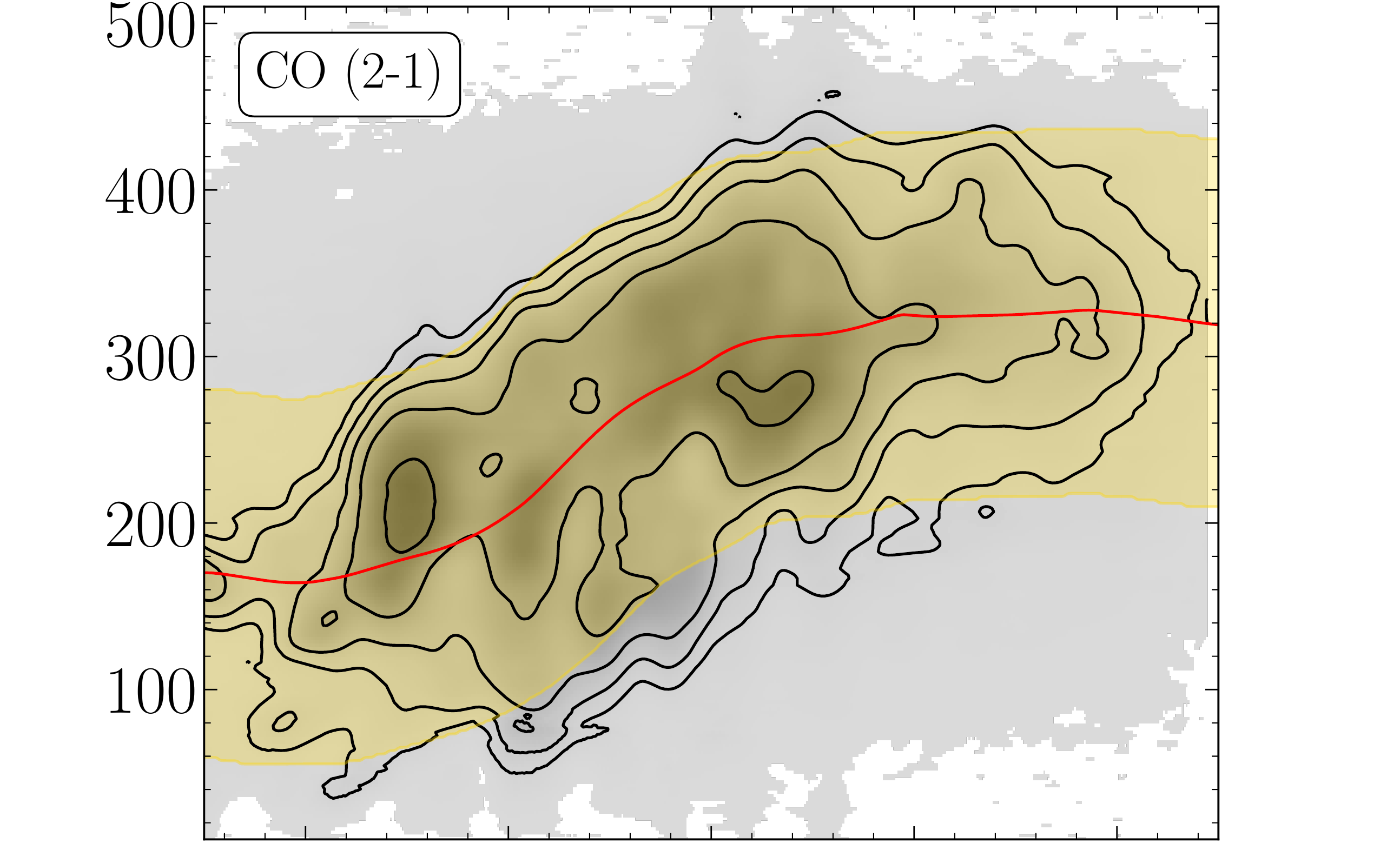}
	\includegraphics[width=\linewidth]{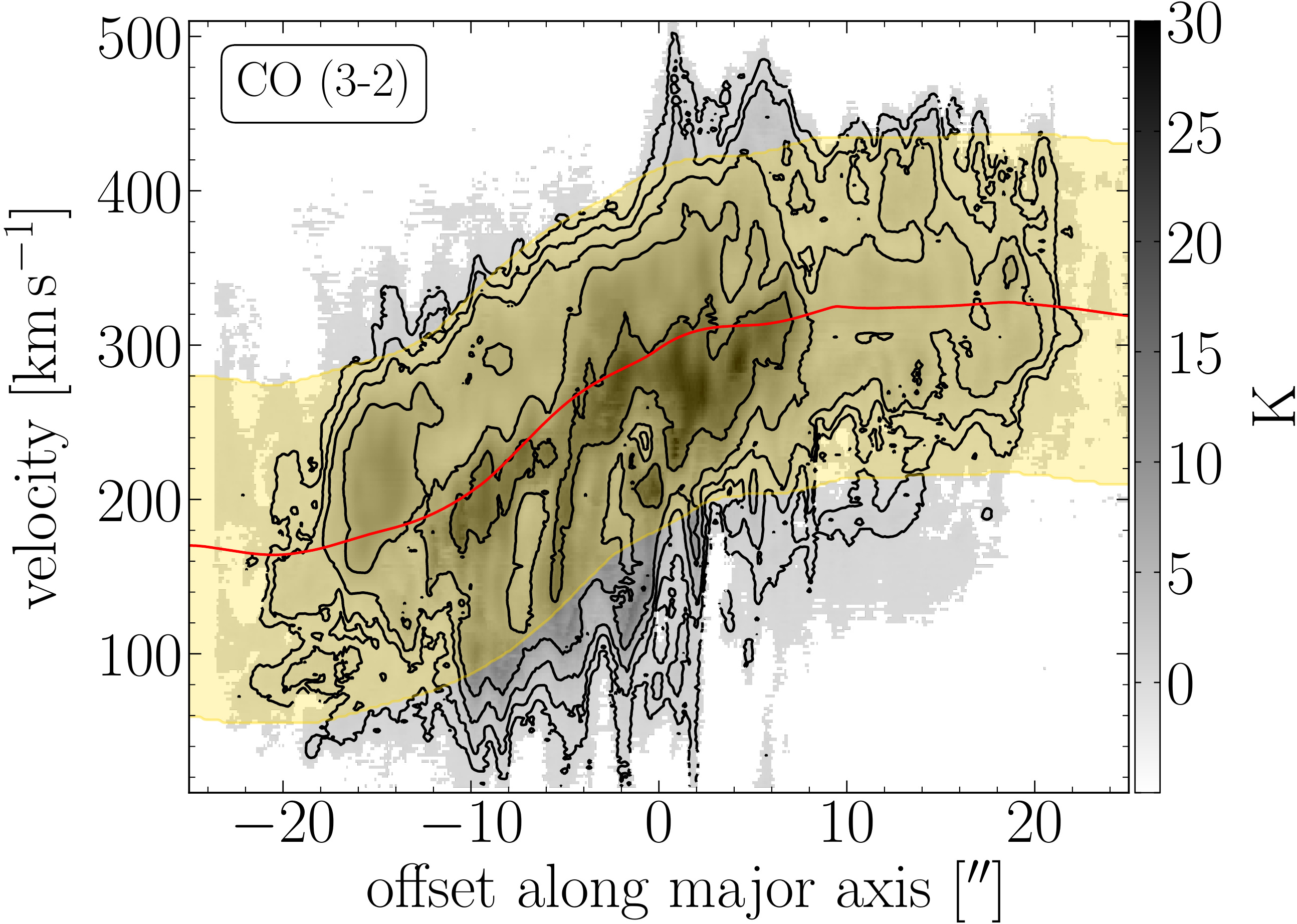}
	\caption{Position-velocity diagram of the central slice (offset $0.0\arcsec$) showing the construction of the disk/non-disk masks. The background images show the flux density above $5.0 \sigma$ for \co10 (top), \co21 (center) and \co32 (bottom) on identical gray scale. Contours are drawn at 1, 2, 4, 8 and 16\,K. The central velocity for our model of the disk emission is illustrated by the red line. The golden-shaded area denotes the disk mask. Similar figures for other offsets are shown in Appendix~\ref{appendix: all pVs}. Note that the different transitions have different angular resolution.}
	\label{figure: sample slice}
\end{figure}

\subsection{Identifying  outflows in the non-disk component}
We identify three different types of structures in the non-disk component (figure~\ref{figure: separated moments}, contours in figure~\ref{figure: nondisk zoomin} highlight these structures): 

(1) Emission that is co-located with the central disk and bar in projection. This is visible as a ridge in \co32 (inner contour in figure~\ref{figure: nondisk zoomin}), and also present but less apparent in \co10 and \co21. The structure is unlikely to be an outflow. It appears more likely to be an additional kinematic component of the disk/bar that is not included in the model we used for the separation. We therefore do not consider this gas to contribute to the total mass outflow rate. 

(2) Emission associated with the so-called western superbubble, located to the west and north of the central starburst region \citep[][shown by the western contour in figure~\ref{figure: nondisk zoomin}]{2006ApJ...636..685S,2013Natur.499..450B}. This feature is already known to be kinematically distinct from the surrounding gas. Part of it is likely the base of the northern outflow cone (and giving rise to the NW streamers identified by \citeauthor{2013Natur.499..450B}, for example), but it is difficult to know what portion of the emission should be associated with a net outflow. 
In our calculations below we exclude this feature from the total outflow rate of NGC\,253, although it likely has some contribution to outflow.

(3) The remaining gas associated with the non-disk component is organized in small clumps along the edge of the disk region or beyond it. Some of this gas is not discernible as individual structures, particularly in the \co10 and \co21 cubes, perhaps due to the resolution but maybe also due to the excitation conditions, constituting extended regions with diffuse emission. Some of the emission is located in well-defined structures known to be part of the outflow, such as the SW streamer which is apparent in all CO transitions.

In summary, the non-disk component consists of these three sub-components: structures that we associate with a net ``outflow,'' structures that are part of the ``western superbubble,'' and structures that are co-located with the ``disk''. The latter is not associated with the outflow, while parts of the western supperbubble may contribute to it. Below we calculate properties for the two components disk and non-disk, and its sub-components individually where it is feasible to do so.


\section{Results}\label{section: results separated disk/non-disk}


The process described in the previous section allows us to estimate the properties of the galactic outflow and other structures. 
A 2D representation of the separated data cubes is shown in figure~\ref{figure: separated moments} in the form of moment maps for integrated intensity (moment 0) and intensity-weighted velocity (moment 1) for all three CO transitions. Striping artifacts due to the pV cuts used in the separation method are present in the disk and non-disk components, visible as straight lines parallel to the major axis. This is primarily aesthetic. We tested their effects on the fluxes and derived velocities by varying the slice width and found them to be negligible.

\begin{figure*}
	\caption{Comparison between original moment maps and separated disk/non-disk components. The outline of the maps is defined by the observed field of view and the square region considered for separating the kinematic components.).
	}
	\subfloat[
	Moment 0 (integrated intensity) of the original image (\emph{top}), disk component (\emph{middle}) and non-disk component (\emph{bottom}). The logarithmic color scales are identical for all panels and chosen to also show the fainter non-disk component which saturates the inner regions of the disk. Contours are drawn at $\log \left( F\ \lbrack \mathrm{K\,km\,s}^{-1} \rbrack \right) = 1.7, 2.0, 2.3, 2.6, 2.9, 3.2, 3.5$; for clarity, only every other contour is drawn for \co32.
	]{
		\centering
		\includegraphics[width=\linewidth]{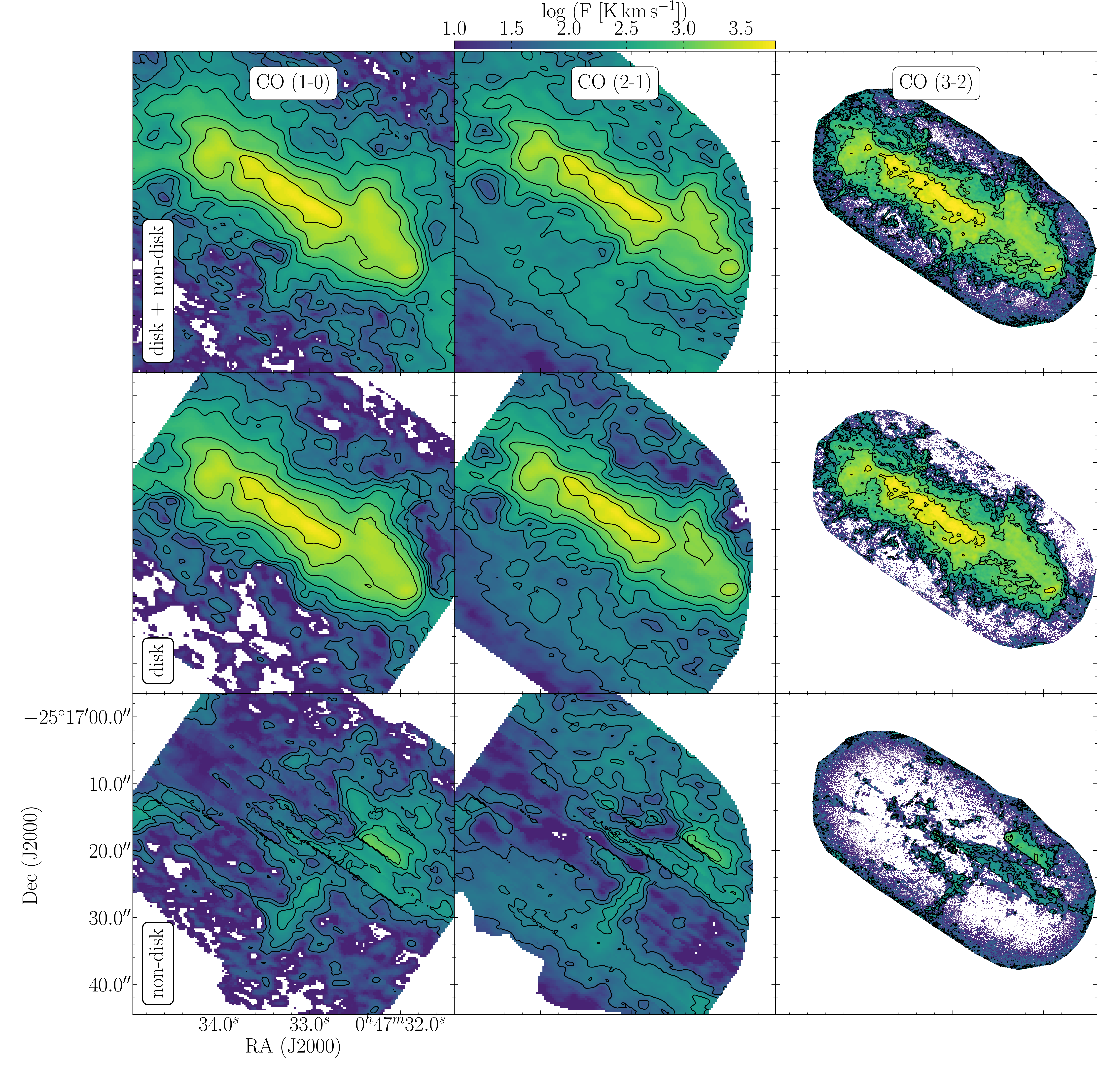}
	}
	\label{figure: separated moments}
\end{figure*}
\begin{figure*}
	\ContinuedFloat
	\centering
	\subfloat[
	Moment 1 (intensity-weighted velocity) of the original image (\emph{top}), disk component (\emph{middle}) and non-disk component (\emph{bottom}). Contours are drawn at 150, 200, ..., 350\,\kms. The noise edge visible in \co32 is due to primary beam correction required to derive accurate fluxes.
	]{
		\centering
		\includegraphics[width=\linewidth]{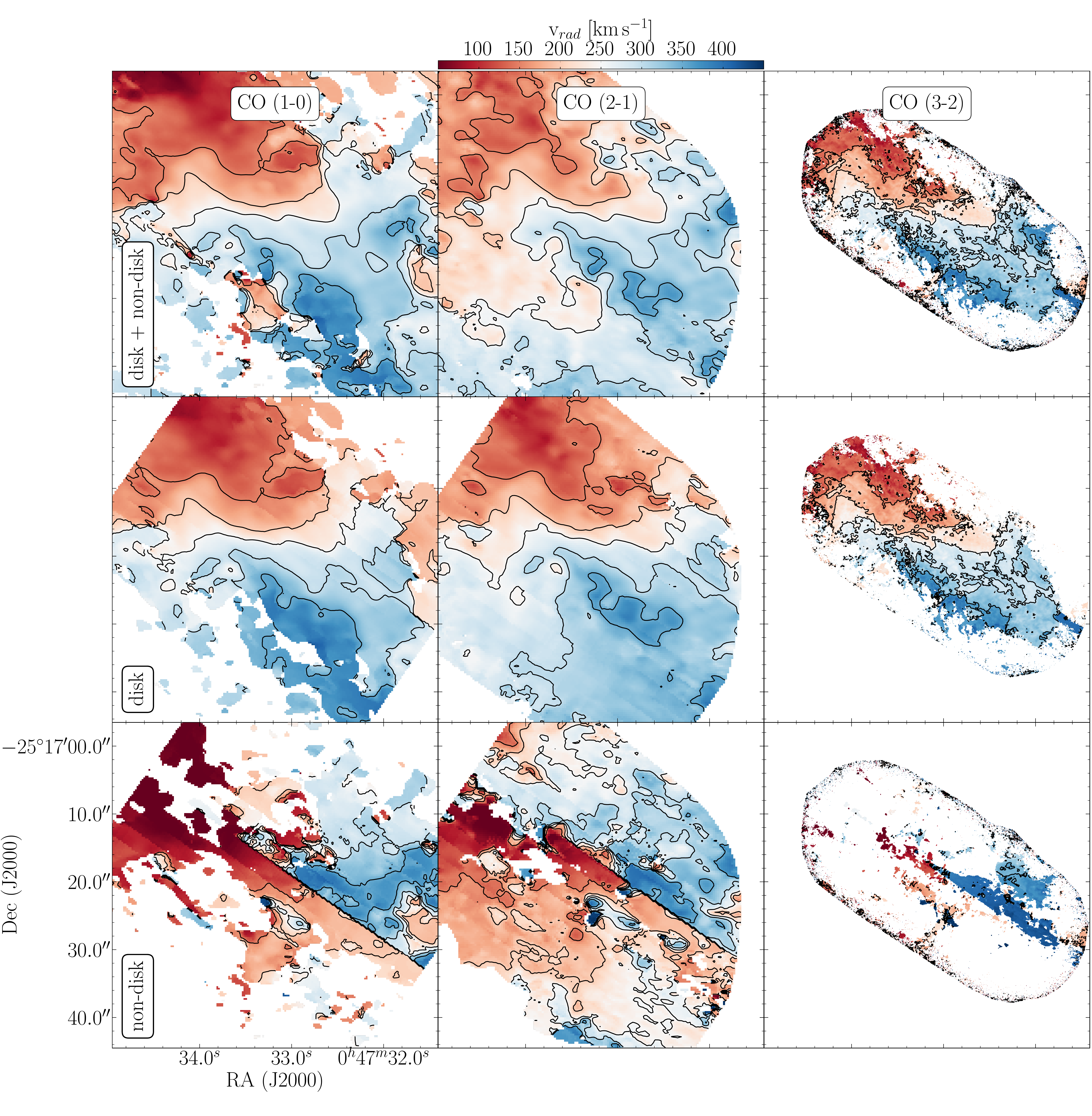}
	}
	\stepcounter{figure}
\end{figure*}

\begin{figure}
    \centering
    \includegraphics[width=\linewidth]{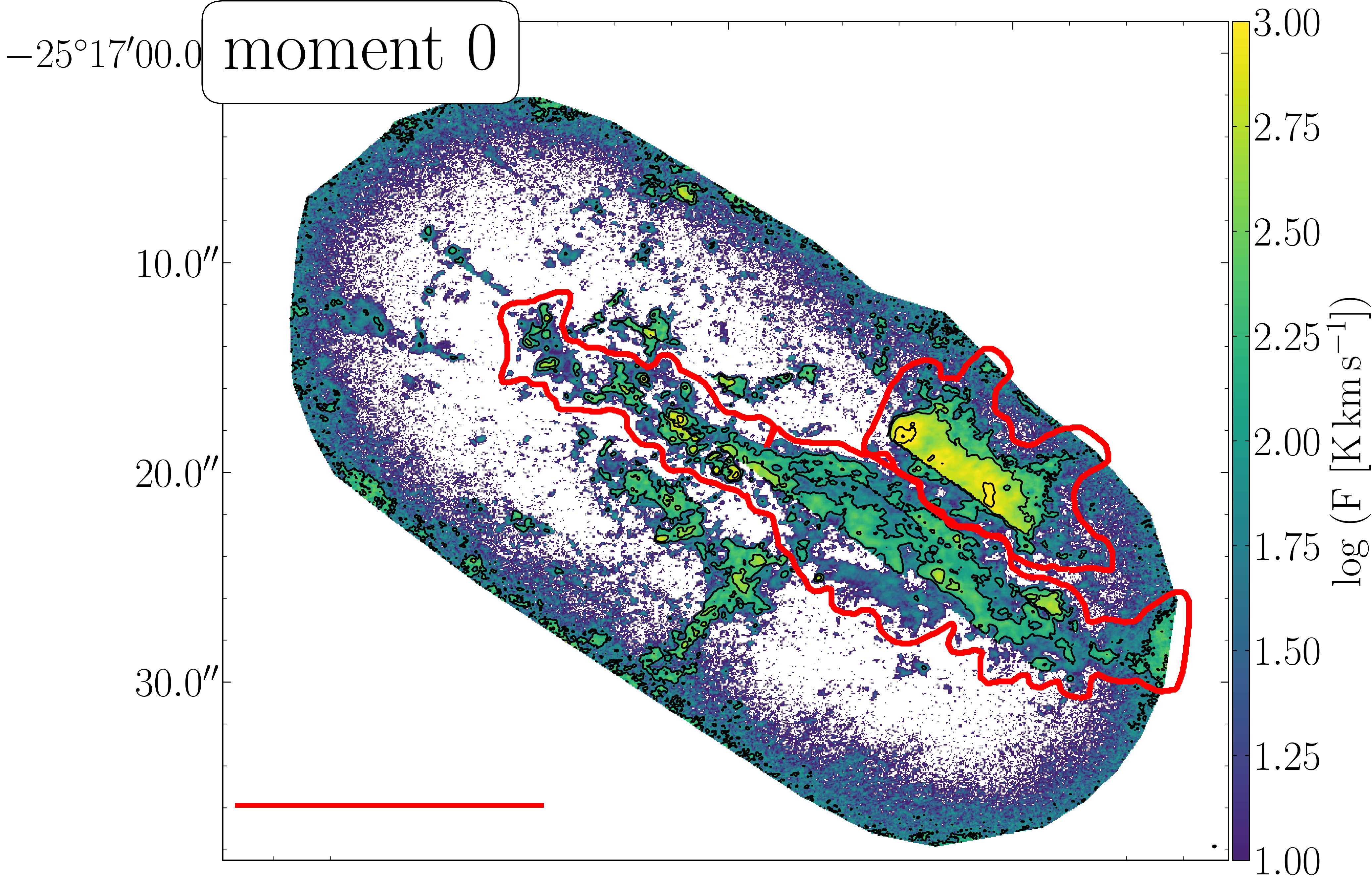}
    \includegraphics[width=\linewidth]{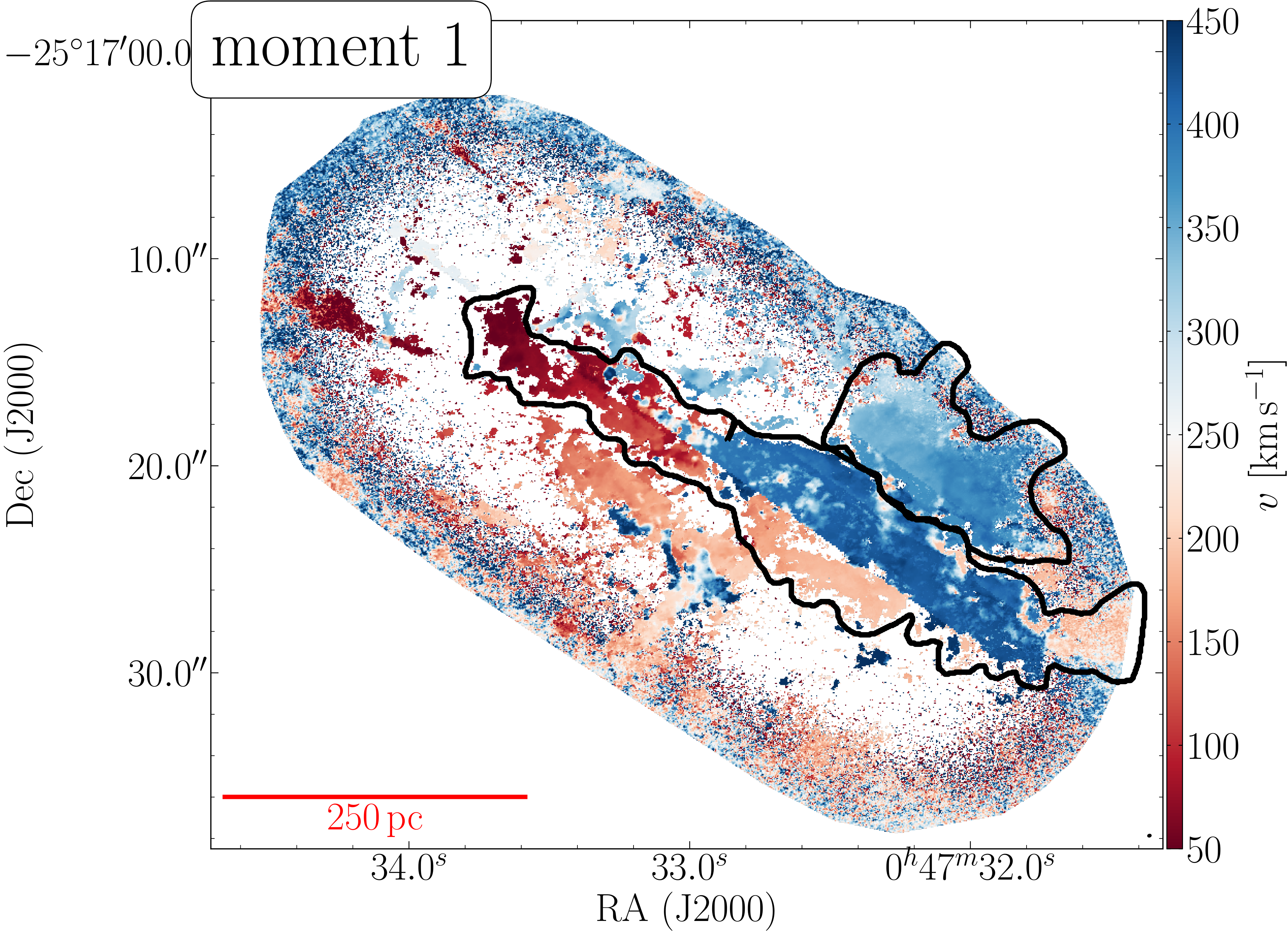}
    \caption{A zoom-in on the non-disk component of \co32 (the bottom left panels in figure~\ref{figure: separated moments}a and \ref{figure: separated moments}b). \emph{top}: Moment~0 (integrated intensity) with contours at $\log \left( F\ \lbrack \mathrm{km\,s}^{-1} \rbrack \right) = 2.0, 2.5, 3.0$. \emph{bottom}: Moment~1 (intensity-weighted velocity). The thick contours show the regions discussed in the text (section~\ref{section: outflow rate}): gas that is kinematically not consistent with disk rotation but co-spatial with the disk in projection, and the western superbubble to the north-west of the disk.}
    \label{figure: nondisk zoomin}
\end{figure}


\floattable
\begin{deluxetable*}{lccccccc}
	\tablewidth{\linewidth}
	\tablecaption{Results of separating disk from non-disk emission in NGC\,253. Uncertainties for these quantities are discussed in the corresponding subsections of section~\ref{section: results separated disk/non-disk}.
	\label{table: results}}
	\tablehead{
		\colhead{quantity} & \colhead{unit} & \multicolumn{2}{c}{\co10} & \multicolumn{2}{c}{\co21} & \multicolumn{2}{c}{\co32}\\
		\colhead{} & \colhead{} & \colhead{disk} & \colhead{non-disk} & \colhead{disk} & \colhead{non-disk} & \colhead{disk} & \colhead{non-disk}
	}
	\startdata
	    \multicolumn{2}{l}{\bf luminosity}\\
        L$_\mathrm{CO}$ & \Kkmspc & $2.8 \times 10^8$ & $4.2 \times 10^7$ & $2.3 \times 10^8$ & $4.5 \times 10^7$ & $1.8 \times 10^8$ & $1.2 \times 10^7$ \\
        fraction & \% & 87 & 13 & 84 & 16 & 94 & 6.5 \\
        \midrule
        \multicolumn{2}{l}{\bf molecular gas mass $\mathrm{M}_\mathrm{mol}$ \tablenotemark{\dag}}\\
        total\tablenotemark{a} & \Msun & $3.1 \times 10^8$ & $4.5 \times 10^7$ & $3.1 \times 10^8$ & $6.1 \times 10^7$ & $2.9 \times 10^8$ & $2.0 \times 10^7$ \\
        \phantom{---}outflow\tablenotemark{b} & \Msun & & $2.7 \times 10^7$ & & $4.1 \times 10^7$ & & $8.3 \times 10^6$ \\
        \phantom{---}superbubble\tablenotemark{c} & \Msun & & $8.9 \times 10^6$ & & $8.9 \times 10^6$ & & $5.8 \times 10^6$ \\
        \phantom{---}other-disk\tablenotemark{d} & \Msun & & $7.6 \times 10^6$ & & $7.4 \times 10^6$ & & $5.8 \times 10^6$ \\
        \midrule
        \multicolumn{2}{l}{\bf molecular mass outflow rate $\dot{M}$} \tablenotemark{\ddag}\\
        \phantom{---}outflow (continuous)\tablenotemark{e} & \Msunyr & & 14 & & 20 & & 2.7 \\
        \phantom{---}outflow (constant)\tablenotemark{f} & \Msunyr & & 29 & & 39 & & 4.8 \\
        \midrule
        \multicolumn{2}{l}{\bf kinetic energy $\mathrm{E}_\mathrm{kin}$ \tablenotemark{\S}}\\
        \phantom{---}outflow (continuous)\tablenotemark{e} & \ergs & & $3.9 \times 10^{54}$ & & $4.5 \times 10^{54}$ & & $6.5 \times 10^{53}$ \\
        \phantom{---}outflow (constant)\tablenotemark{f} & \ergs & & $2.5 \times 10^{54}$ & & $3.1 \times 10^{54}$ & & $4.3 \times 10^{53}$ \\
        \midrule
        \multicolumn{2}{l}{\bf momentum $\mathrm{P}$ \tablenotemark{\P}}\\
        \phantom{---}outflow (continuous)\tablenotemark{e} & \Msunkms & & $6.9 \times 10^8$ & & $8.7 \times 10^8$ & & $1.2 \times 10^8$ \\
        \phantom{---}outflow (constant)\tablenotemark{f} & \Msunkms & & $4.8 \times 10^8$ & & $6.4 \times 10^8$ & & $8.0 \times 10^7$ \\
    \enddata
    \tablenotetext{a}{CO line luminosity of all emission considered consistent with disk rotation (disk) and not consistent with disk rotation (non-disk), respectively.}
    \tablenotetext{b}{Non-disk excluding the western superbubble and the gas that is co-spatial with the projected disk.}
	\tablenotetext{c}{Non-disk emission belonging to the western superbubble as defined by \citet{2006ApJ...636..685S}}
	\tablenotetext{d}{Non-disk gas that is co-spatial with the disk in projection. See section~\ref{section: outflow rate} for the definition.}
	\tablenotetext{e}{Outflowing gas as defined by note $^b$ under the assumption of continuous mass ejection without accelerations to the gas after ejection.}
	\tablenotetext{f}{Outflowing gas as defined by note $^b$ under the assumption of approximately constant starting mass outflow rate over the lifetime of the starburst.}
	\tablenotetext{\dag}{Molecular gas mass derived using a conversion factor for \co10 emission of $\mathrm{X}_{\mathrm{CO}} = 0.5\times10^{20}\,\left(\mathrm{K\,km\,s}^{-1}\right)^{-1}$, including the contribution of Helium, and assuming CO brightness temperature line ratios of $r_{21} = 0.80$ and $r_{31} = 0.67$ for \co21 and \co32 relative to \co10.}
	\tablenotetext{\ddag}{Deprojected molecular mass outflow rate. $50^{th}$ percentile best estimate assuming a flat distribution of outflow inclinations for the unknown geometry.}
	\tablenotetext{\S}{Deprojected kinetic energy of the molecular gas. $50^{th}$ percentile best estimate assuming a flat distribution of outflow inclinations for the unknown geometry.}
	\tablenotetext{\P}{Deprojected momentum of the molecular gas. $50^{th}$ percentile best estimate assuming a flat distribution of outflow inclinations for the unknown geometry.}
	\tablecomments{Sources of error are discussed and quantified in the respective subsections of section \ref{section: results separated disk/non-disk}.}
\end{deluxetable*}


\subsection{CO luminosities}

We quantify in table~\ref{table: results} the CO luminosities of the disk and non-disk components. We measure luminosities of $2.8 \times 10^8$\,\Kkmspc, $2.3 \times 10^8$\,\Kkmspc and $1.8 \times 10^8$\,\Kkmspc for \co10, (2--1) and (3--2), respectively, in the central disk of NGC\,253. The non-disk component is, naturally, much fainter with luminosities of $\sim 4.2 \times 10^7$\,\Kkmspc for \co10, $\sim 4.2 \times 10^7$\,\Kkmspc for (2--1) and $\sim 4.2 \times 10^7$\,\Kkmspc (3--2). These correspond to approximately $12.9\%$, 16.4\% and 6.5\% of the total luminosity. These luminosities are measured over the sliced area (cf.\ figure~\ref{figure: slice positions}) for which the coverage is not the same among the datasets. We therefore also measure luminosities integrated over the same spatial region, here defined as the overlap between the datasets. This overlap amounts to 885\,\arcsec$^2$ ($2.55\times10^5$\,pc$^2$). The luminosities in the overlap area are: disk: $2.6 \times 10^8$\,\Kkmspc, $2.1 \times 10^8$\,\Kkmspc and $1.8 \times 10^8$\,\Kkmspc; non-disk: $2.6 \times 10^7$\,\Kkmspc, $2.4 \times 10^7$\,\Kkmspc and $1.2 \times 10^7$\,\Kkmspc for \co10, (2--1) and (3--2), respectively.

An interesting result coming out of our decomposition is that not all the material we identify as ``outflow'' is in well-defined structures such as the streamers identified by \citet{2013Natur.499..450B}. Correctly estimating the outflow rate requires accounting also for a diffuse extended component. 

It is important to compare our fluxes to measurements in the literature. \citet{1996A&A...305..421M} find a \co21 luminosity of $1.2 \times 10^6$\,K\,\kms\,arcsec$^2$ which translates\footnote{adjusting from the distance $\mathrm{D} = 2.5$\,pc used by \citeauthor{1996A&A...305..421M} to the $\mathrm{D} = 3.5$\,pc assumed here} to $3.5 \times 10^8$\,\Kkmspc or $1.3$ times our measurement. Their observations cover $80\arcsec \times 60\arcsec$,
a area similar to our \co10 observations (but $\sim 4$ times larger than the area of our \co32 observations). For the outflow \co10 luminosity, \citet{2013Natur.499..450B} derive an estimate of $2.0 \times 10^7$\,\Kkmspc by summing over individual identified molecular outflow features. This includes flux from the ``superbubble'' component, so it is probably better compared to the sum of our ``outflow'' and ``superbubble'' components of $\sim3.6\times10^7$\,\Kkmspc. Given the large methodological differences and the importance of the diffuse emission, these numbers are in reasonable agreement.


\subsection{Masses of components}\label{section: mass distribution}

The total gas mass $M$ is estimated from the CO line luminosity, using the conversion factor $\mathrm{X}_{\mathrm{CO}} = 0.5\times10^{20}\,\left(\mathrm{K\,km\,s}^{-1}\right)^{-1}$\,\sqcm corresponding to $\alpha_\mathrm{CO} = 1.1\,\mathrm{M}_\odot\,\left(\mathrm{K\,km\,s}^{-1}\,\mathrm{pc}^{-1}\right)^{-1}$ discussed by \citet{Leroy:2015ds} for the central starburst region. This value accounts for the effects of moderate optical depth, high velocity dispersion, and warm gas temperatures that are likely to dominate the central regions of NGC\,253. The masses we report include the contribution of Helium to the total mass. To compute masses using the \co21 and \co32 transitions we assume typical line ratios of $r_{21} = 0.80$ and $r_{31} = 0.67$ relative to \co10 as implied by \citet{2018ApJ...867..111Z}. Note that we do not measure line ratios from the images but adopt a uniform factor to keep the mass measurements from the three observed CO lines independent.

Table~\ref{table: results} lists the masses corresponding to the disk and the non-disk components. Uncertainty in the mass estimates arises primarily from the assumed conversion factor and the apportioning of emission among the different components. The calibration uncertainty for the flux measurements is $\sim10-15\%$ for the ALMA observations. Overall, we adopt a systematic error of factor $\sim2$ for the the derived masses.

The molecular masses derived from the three CO transitions are very similar. They match within 10\% for the disk component, and within 50\% for the non-disk component. We estimate the total gas mass in the center of NGC\,253 to be $\sim 3.5 \times 10^8$\,M$_\odot$ (adding the disk and non-disk components), with estimates in the range of $3.1-3.6 \times 10^8$\,\Msun for the different transitions. About 85\% of the total mass is in the disk component. The masses estimated in the non-disk components using \co10 and \co21 are fairly similar at $4.5 \times 10^7$\,\Msun and $6.1 \times 10^7$\,\Msun whereas in \co32 we detect a lower $2.0 \times 10^7$\,\Msun, a consequence of the lower luminosity. The non-disk masses are primarily contributed by the outflow component ($\sim 50$\%). About $20-30$\% of mass is in the western supperbubble and $12-30$\% is co-spatial with the disk but kinematically distinct.

It is important to compare these mass estimates to previous results for the total molecular gas mass in NGC\,253, noting that our analysis
covers the central $45\arcsec \times 25\arcsec$ ($750\,\mathrm{pc} \times 400\,\mathrm{pc}$). Towards the east, $\sim 10\%$ of the known molecular gas close to the center is not covered by our \co32 observations and thus not considered in this analysis. The agreement with previous measurements is very good. \citet{1996A&A...305..421M} reported a mass of $1.3 \times 10^8$\,\Msun over a similar area ($80\arcsec \times 60\arcsec$ in the center of NGC\,253), but this was based on a different distance and $\mathrm{X}_{\mathrm{CO}}$. After correcting for those differences, their luminosity corresponds to $4.2 \times 10^8$\,\Msun, consistent with our measurement. Using the same distance and the same 1--0 observations, \citet{Leroy:2015ds} measure a molecular mass of $3.5 \times 10^8$\,\Msun. 
\citet{2018ApJ...860...23P} report a total gas mass of $4.5 \times 10^8$\,\Msun\ derived from the sub-mm dust spectral energy distribution, which is very consistent with our result given the very different methodologies. 

No estimates in the literature separate the ``disk'' and ``non-disk'' components as we do above. Previous estimates of the outflowing mass range from a lower limit of $6.6 \times 10^6$\,\Msun calculated for the optically thin limit \citep{2013Natur.499..450B}, to $2-4 \times 10^7$\,\Msun  when accounting for optical depth \citep{2018ApJ...867..111Z}. Since we identify an outflowing mass $\sim5\times10^7$\,\Msun, the agreement with the latter estimate is fairly good. Note, however, that these studies derive the outflowing mass from individual features rather than using the position-velocity information as we do here in a systematic way.


\subsection{Mass outflow rate}\label{section: outflow rate}

\begin{figure}
    \centering
    \includegraphics[width=\linewidth]{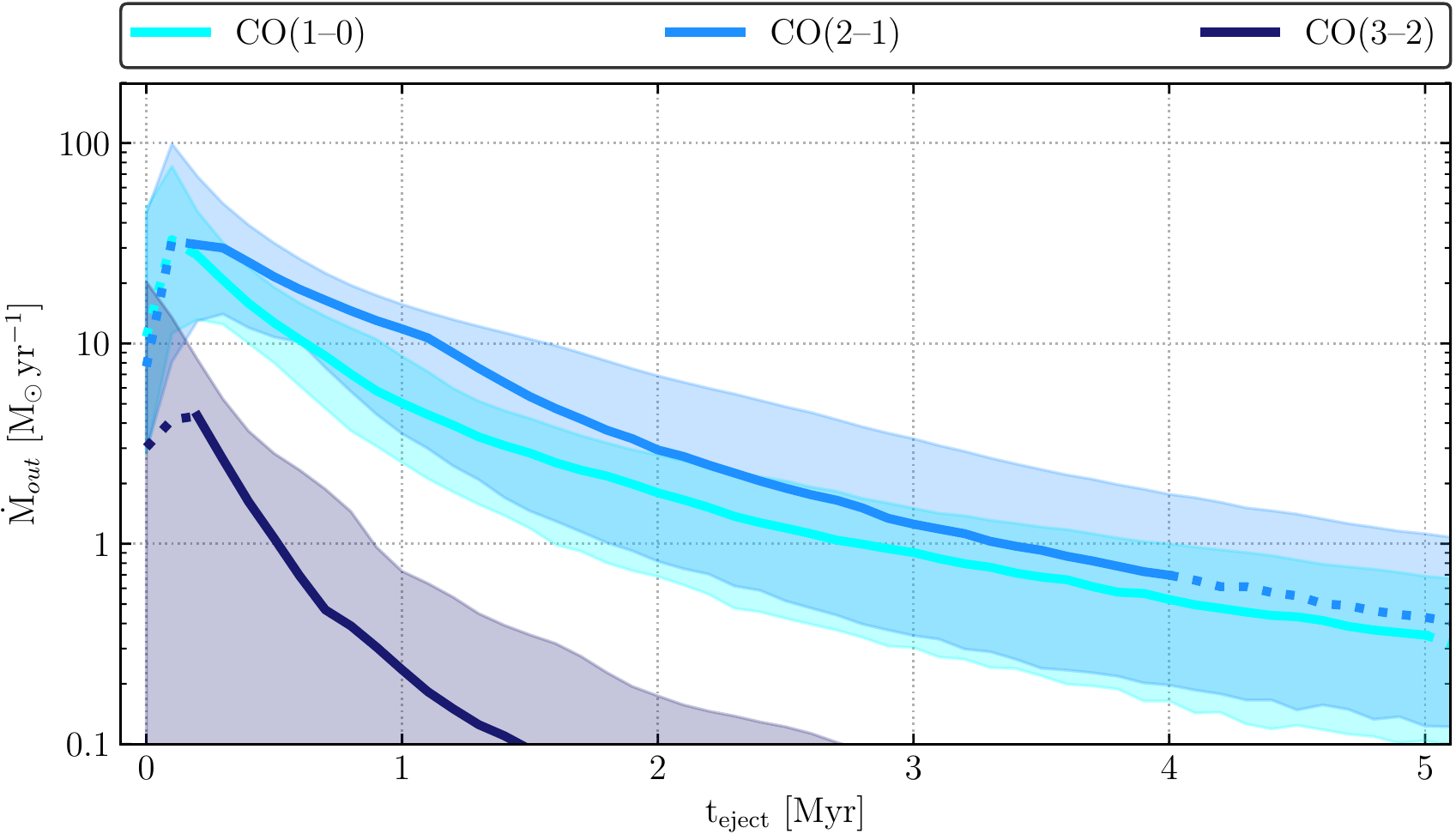}
    \\ \vspace{0.2cm}
    \includegraphics[width=\linewidth]{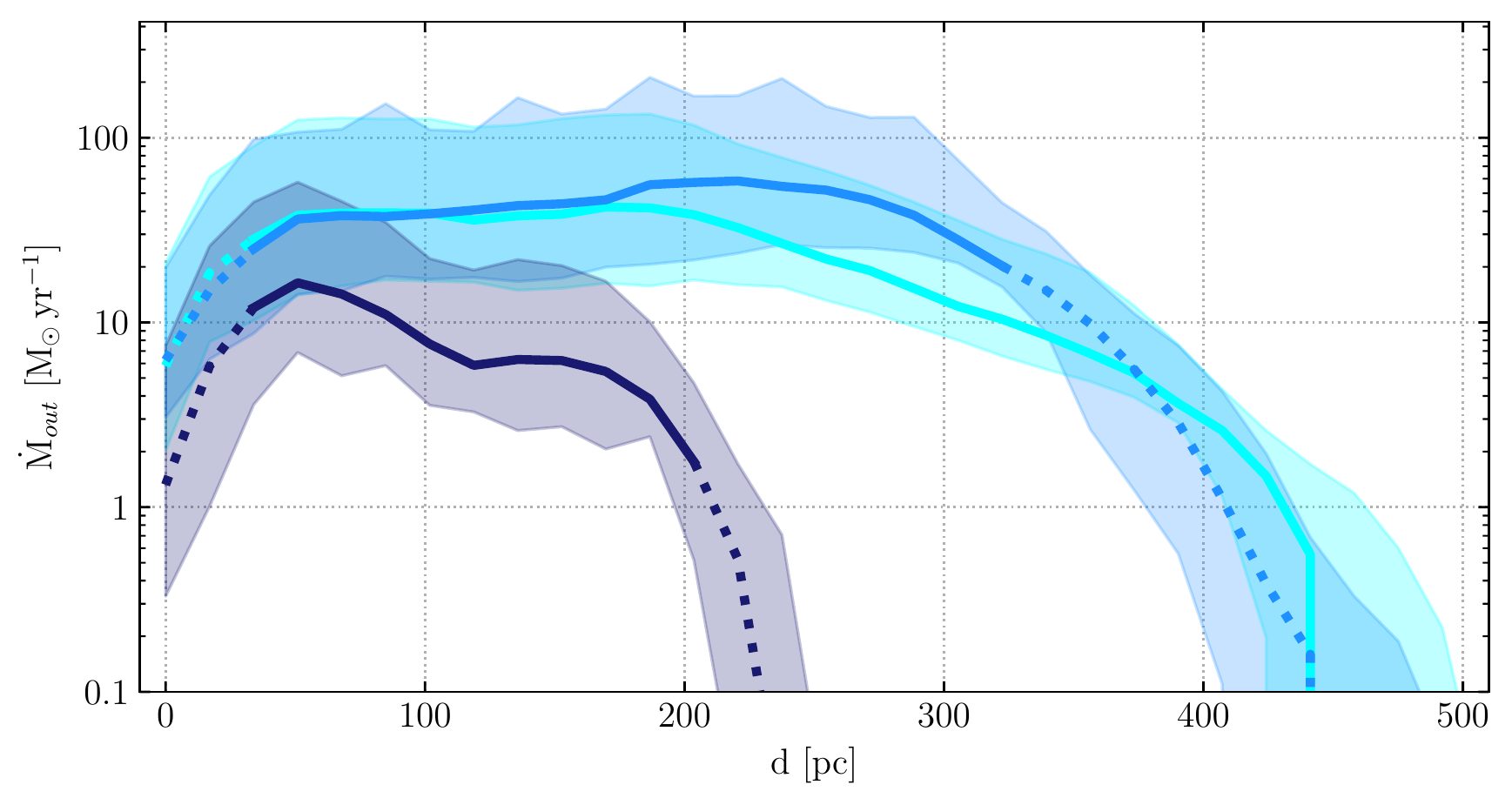}
    \caption{Deprojected molecular mass outflow rate averaged over 0.1\,Myr as a function of time since ejection (\emph{top}) and as a function of deprojected distance between outflow and launching site (\emph{bottom}). The top panel implicitly assumes continuous mass ejection without accelerations to the gas after ejection,  while the lower panel assumes approximately constant starting mass outflow rate over the lifetime of the starburst. The shaded area indicates approximate errors ($16^{th}$ to $84^{th}$ percentile), which are dominated by uncertainties in the deprojection geometry. Dotted lines represent the ranges where confusion with gas in the disk occurs and where the limited field-of-view affects the completeness.}
    \label{figure: outflow rate}
\end{figure}

A mass flow rate is defined as the flux of mass per unit time through a surface. In our case, we are interested in the flow of molecular gas mass through a virtual closed surface around the center of NGC\,253 at a given distance. Note that an individual outflow feature observed over a certain length, such as the SW streamer, can develop in at least two ways: as the distance of an outflow from its origin corresponds to time since ejection times velocity, continuously outflowing gas results in extended streaming structures. Gas ejected at a single ejection event in the past with a distribution of ejection velocities, on the other hand, will also result in an extended streamer. In reality gas will be ejected with a distribution of velocities at a varying rate over a period of time, and in order to interpret the measurements we need to make some simplifying assumptions. We chose two edge cases to span the range of different interpretations: (1) the gas does not experience accelerations after being launched \citep[however, see ][]{2017ApJ...835..265W}, and (2) that the gas outflow rate is approximately constant with time. For all calculations, however, we assume that the projected direction of flow is perpendicular the the central plane of the bar and that CO emission traces the mass with a constant conversion factor.

We compute both the outflow rate as a function of distance and as a function of time. If we assume that the mass outflow rate has been approximately constant over the lifetime of the starburst, for example, a diminishing outflow rate as a function of distance would suggest that gas is either launched with or somehow develops a distribution of velocities. Conversely, if we assume that the present day velocity has been constant since the gas was ejected, we can derive a history of the mass outflow rate as a function of time and account for a variable mass outflow rate. Both interpretations are equally, but not simultaneously, valid.

For the detailed calculation of the mass outflow rate, we proceed as follows. A mass outflow rate is $\dot{\mathrm{M}} = \mathrm{M} \,t^{-1}$ with mass M and relevant time scale $t$. For each image element $i$ (3D pixel or sometimes also called voxel), we calculate the outflow rate $\dot{m}_i$ of the gas that was ejected at time $t_{{\mathrm{eject}},i}$ over the time interval $\Delta t_{\mathrm{cross},i}$, as the ratio of mass of the pixel ${m}_i$ to the pixel crossing interval $t_{\mathrm{cross},i}$. Ejection time and pixel crossing interval are functions of the outflow velocity $v_i$ and the distance $s_i$ between current pixel position and the launching site, and the pixel size in the direction of the flow $\Delta s$, respectively. We therefore compute
\begin{eqnarray}
    \Delta t_{\mathrm{cross},i}   &=& \frac{\Delta s}{v_i}\\    
    \dot{m}_i\ (t_{\mathrm{eject},i}) &=& \frac{m_i}{\Delta t_{\mathrm{cross},i}}\\
    t_{\mathrm{eject},i} &=& \frac{s_i}{v_i}
\end{eqnarray}

\noindent obtaining a mass outflow rate $\dot{m}_i$, a distance $s_i$, and an ejection time $t_{\mathrm{eject},i}$ for each pixel in the ``outflow'' component. Note that this approach takes the 3D phase space information into account by treating pixels independently. Typically, a sightline shows multiple pixels with emission at different velocities that all contribute an outflow rate with their respective mass, distance and velocity. We then bin the outflow rates $\dot{m}_i$ by ejection time $t_{\mathrm{eject},i}$ and integrate over the time range $\left[\mathrm{T}_1, \mathrm{T}_2 \right]$ to obtain the average outflow rate in this time interval,
\begin{equation}
    \dot{M} \left( \mathrm{T}_1, \mathrm{T}_2 \right) = 
    \frac{\displaystyle \sum_i \dot{m}_i \left( \mathrm{T}_1 < t_{\mathrm{eject},i} < \mathrm{T}_2 \right)  \Delta t_{\mathrm{cross},i} }{\displaystyle \mathrm{T}_2 - \mathrm{T}_1 }.
    \label{equation: outflow rate time}
\end{equation}

\noindent Similarly, binning by distance results in the average outflow rate at a given distance, 
\begin{equation}
    \dot{M} \left( \mathrm{D}_1, \mathrm{D}_2 \right) =
    \frac{\displaystyle \sum_i \dot{m}_i \left( \mathrm{D}_1 < s_i < \mathrm{D}_2 \right) \Delta s }{\displaystyle \mathrm{D}_2 - \mathrm{D}_1 }.
    \label{equation: outflow rate distance}
\end{equation}
\noindent Performing binning on a sequence of time intervals yields the outflow rate history, while binning in distance tells us how far from the launching site a given fraction of the mass is able to escape.

Calculating velocity $v$ and distance $s$ requires knowledge about the geometry and origin of each outflowing gas parcel. The simplest assumption, used here, is that on average outflows are launched in the plane of the central region of the galaxy which corresponds to launching on the major axis. The distance $s$ is thus the projected distance to the major axis on the edge of an outflow cone with given opening angle. Note that the outflow originates from an extended region in the disk and the term cone thus refers to a cut-off cone (called a frustum in geometry). Velocity $v$ is the velocity difference between launching site and current velocity of the outflow parcel, i.e. the velocity difference over distance $s$. Both the velocity of the launching site and the projected distance are uncertain. The velocity changes by $\pm 25$\,\kms when an outflow originates from the northern/southern edge of the observed CO disk (above/below the plane), while the projected distance traveled by the gas changes by $\pm1.25\arcsec$ ($\pm20$\,pc).

Distance $s$, ejection time scale $t_\mathrm{eject}$ and pixel crossing time scale $t_\mathrm{cross}$ are measured as projected quantities that need to be deprojected to account for the outflow geometry. The bright molecular streamers (the SW and SE streamers) seem to lie at the edge of the ionized outflow cone with $\sim60^\circ$ opening angle \citep{2013Natur.499..450B}. Assuming that all molecular outflows are along this cone, and that the axis of the cone is oriented perpendicular to the disk ($i = 78^\circ$), the range of effective inclination of outflowing gas can be anywhere between $\theta = 48^\circ$ and $\theta = 108^\circ$. Deprojected velocity, $v_\mathrm{depro} = v_\mathrm{obs} / \sin\theta$, and distance, $s_\mathrm{depro} = s_\mathrm{obs} / \cos\theta$, have a direct effect on the deprojected outflow rate, $\dot{m}_\mathrm{depro} = \dot{m}_\mathrm{obs} \tan\theta$, and also on the inferred time and distance evolution of the outflow rate. We use a Monte Carlo approach to derive the errors introduced by deprojection, assuming that the outflow direction has an equal probability of being in any direction along the surface of the outflow cone.

Figure~\ref{figure: outflow rate} shows the molecular mass outflow rate as a function of time or distance, corresponding to the two alternative interpretations we discuss above: a flow where the distribution of material is interpreted as resulting from the history of mass outflow rate (top panel), and one where we show the mass outflow rate as a function of distance, which under the assumption of a constant outflow rate over the last several Myr can be interpreted as an efficiency of ejection to a given distance (bottom panel). Indeed, for an outflow with a distribution of velocities the slower material will not travel as far in a given time, neither will it escape the galaxy if it does not have a high enough velocity. 
Close to the starburst region (or at small times since ejection) the mass outflow rate drops to zero, because it becomes increasingly difficult to separate the ``outflow'' component from the ``disk'' component. At large values of distance or time it also drops to zero, due to a decreasing amount of outflowing molecular material detected far from the starbursts (and the fact that the observations have a limited field-of-view).
The constant outflow rate out to $\sim 300$\,pc is in tension with \citet{2018ApJ...853..173K} who find a steeply dropping ``cold'' ($\mathrm{T}<5050$\,K) component in their TIGRESS simulation. Within 400\,pc, they find the averaged mass loading factor to drop by two orders of magnitude. It is unlikely that the SFR in NGC\,253 has increased by two orders of magnitude within the past 1-2\,Gyr which would alter the observed constant outflow rate profile to be consistent with the \citet{2018ApJ...853..173K} simulation. Note, however, that their simulation recreates solar neighborhood-like conditions instead of a starburst. A direct comparison may thus be not possible.

Our data show that the \emph{average} outflow rates within 20\arcsec (340\,pc) from the major axis are 29\,\Msunyr ($^{+0.48}_{-0.35}$\,dex), 39\,\Msunyr ($^{+0.49}_{-0.34}$\,dex) and 4.8\,\Msunyr ($^{+0.50}_{-0.39}$\,dex) for \co10, (2--1) and (3--2) respectively. Similarly, within the past 1.0\,Myr, the \emph{average} outflow rates are 14\,\Msunyr ($^{+0.25}_{-0.29}$\,dex), 20\,\Msunyr ($^{+0.27}_{-0.37}$\,dex) and 2.7\,\Msunyr ($^{+0.22}_{-0.56}$\,dex) for \co10, (2--1) and (3--2) respectively. The uncertainties, indicated by the $16^{th}$ to $84^{th}$ percentile in the Monte Carlo described above, are substantial at a factor of $2-3$. Real systematic uncertainties are even larger, since there can be conversion of molecular into atomic material \citep[c.f. ][]{2015ApJ...814...83L}, or in general variations in the CO-to-H$_2$ conversion.

Note that the average outflow rates quoted above differ between the two representations, with the median outflow rate as a function of distance being about twice as high as a function of time. The outflowing mass is identical in both cases, and the difference arises solely from binning. Comparing between lines, it is apparent that as measured in \co32 the outflow rate is roughly one order of magnitude lower than for the lower two transitions. This is a direct consequence of the lower mass detected in \co32, and the smaller field-of-view of those observations. The \co32 observations cover only $\sim 12.5\arcsec$ ($\sim 210$\,pc projected) above/below the disk and thus miss significant amounts of non-disk gas. Their lower surface brightness sensitivity means we also fail to detect a diffuse non-disk component, as we see in the two lower lines. The measurements in \co32 should thus be interpreted as a lower limit, and in that sense they are consistent with those for the lower two transitions.

Overall, the deprojected total mass outflow rate in the starburst of NGC\,253 is most likely in the range $\sim 14-39$\,\Msunyr as derived from \co10 and \co21 with $\sim 0.4$\,dex uncertainty. The large spread arises due to different interpretations of the kinematics of the observed gas while the errors are due to unknown geometry.
The majority of this outflow rate is contributed by massive outflows alongside the disk like the SW/SE streamers, with a significant contribution by diffuse molecular gas.

The present day star formation rate in the central region of NGC\,253 is  $1.7-2.8$\,\Msunyr, derived from radio continuum and far-infrared measurements \citep{Ott:2005il,Leroy:2015ds,2015MNRAS.450L..80B}. This results in a mass loading factor $\eta = \dot{M}_\mathrm{out} / \dot{M}_\mathrm{SFR}$ in the range $\eta \sim 5.4-23.5$. Note that this is for gas ejected as far as 340\,pc. We do not currently know what fraction of the gas makes it to the far regions of the halo, or reaches escape velocity from the system. Theoretical works suggest that most of the molecular outflow will not escape but rain back down on the galaxy (e.g. \citealt{Shapiro:1976ha} up to recent work by \citealt{2018ApJ...853..173K} or \citealt{2019MNRAS.tmp..540T}).

In our data, no gas reaches the escape velocity of $v_\mathrm{esc} = 500$\,\kms \citep{2017ApJ...835..265W}. The uncertainty on $v_\mathrm{esc}$ is substantial, so allowing a factor of two is still plausible. At $v_\mathrm{esc} = 250$\,\kms, the fraction of gas above $v_\mathrm{esc}$ by mass is 0.5\%, 0.5\% and 6.0\% for \co10, \co21 and \co32, respectively. The mismatch between the lower transitions and \co32 implies that some high velocity gas can be found on small scales that is blurred out in the low resolution observations.

This estimate of the molecular mass outflow rate is higher than the lower limit found by \citet{2013Natur.499..450B} for optically thin emission. \citet{2018ApJ...867..111Z} analysis of the CO line ratios in the SW streamer shows that the emission there is optically thick, which the authors used this to rescale the \citet{2013Natur.499..450B} measurements finding a NGC\,253 galactic outflow rate of $25-50$\,\Msunyr. The result presented here, a mass outflow rate of $\sim 14-39$\,\Msunyr, is consistent with this number using an independent and a more complete methodology than the original work.

From H$\alpha$ observations by \citet{Westmoquette:2011bp}, we can estimate the ionized outflow rate to $\sim 4$\,\Msunyr using their ionized mass ($\mathrm{M} = 10^7$\,\Msun) and typical velocity (200\,\kms) at mean deprojected distance (510\,pc). X-ray observations yield comparable values. \citet{Strickland:2000wd} finds an upper limit of 2.2\,\Msunyr assuming a standard outflow velocity of 3000\,\kms. The upper limit reported in \citet{Strickland:2002kp} translates to 2.3\,\Msunyr when assuming 3000\,\kms outflow velocity and a reasonable 10\% filling factor. These estimates scale linearly with the unknown velocity and also depend on the unknown metallicity and filling factor in the outflow. Estimates of the outflow rate in neutral gas are not known in the literature but are arguably at a similar level.
The molecular phase thus clearly dominates the mass budget in the outflow close to the disk as found in other galaxies (e.g. M82, \citealt{2015ApJ...814...83L} and simulations, e.g. \citealt{2018ApJ...853..173K}).


\subsection{Outflow energy and momentum}\label{section: outflow energy}

\begin{figure*}
    \centering
    \includegraphics[width=0.48\linewidth]{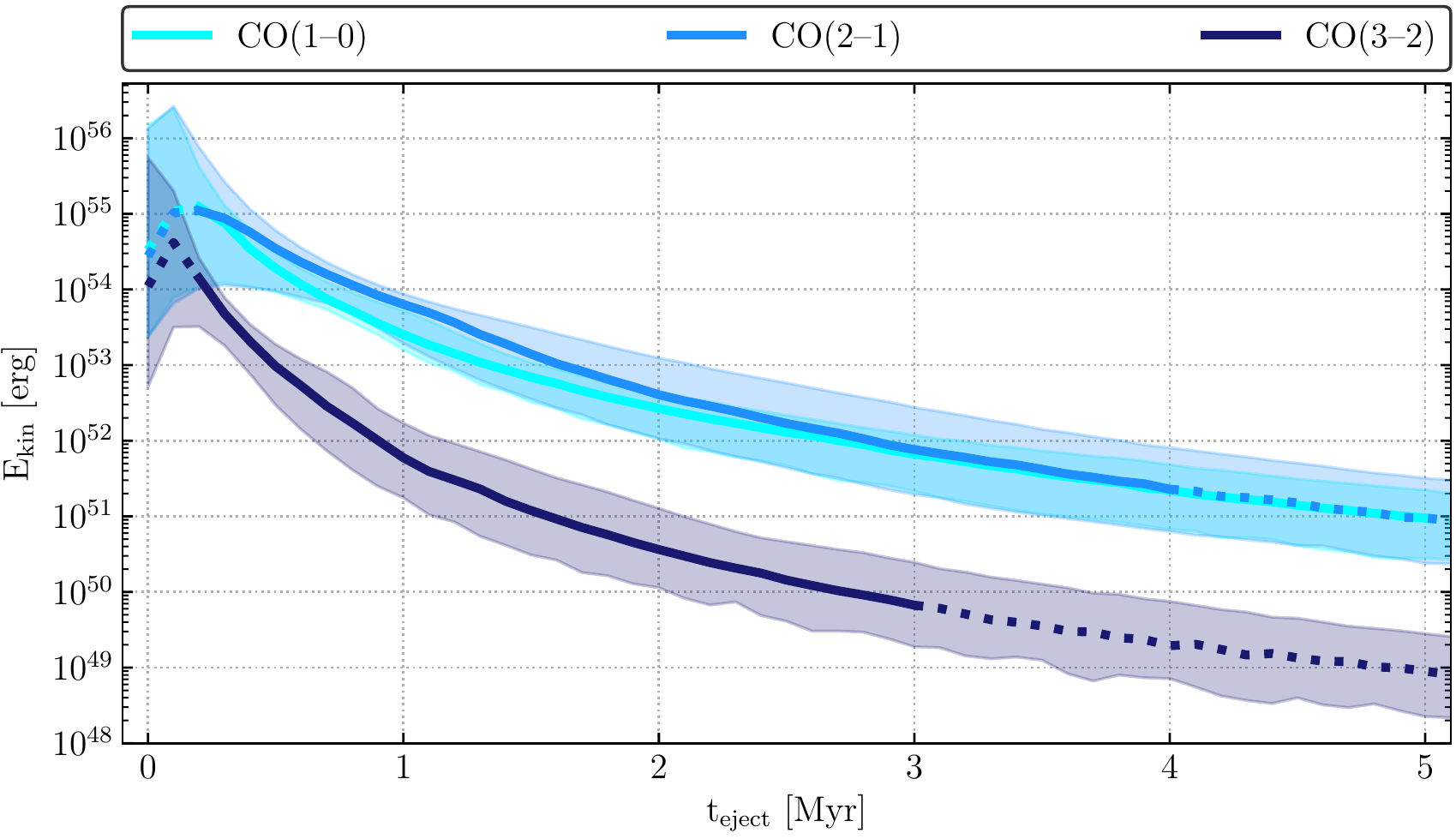}~~
    \includegraphics[width=0.48\linewidth]{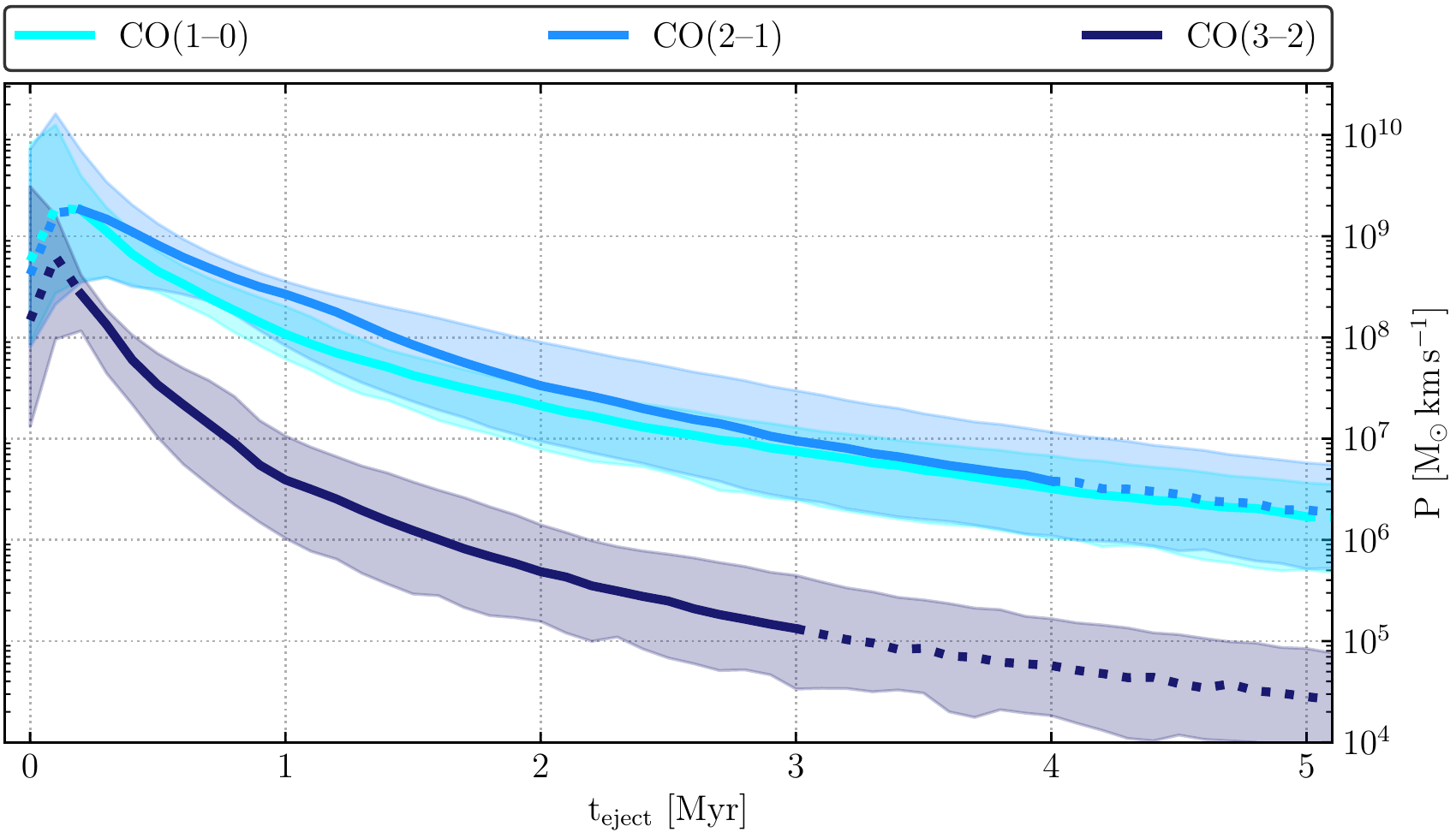}
    \\
    \includegraphics[width=0.48\linewidth]{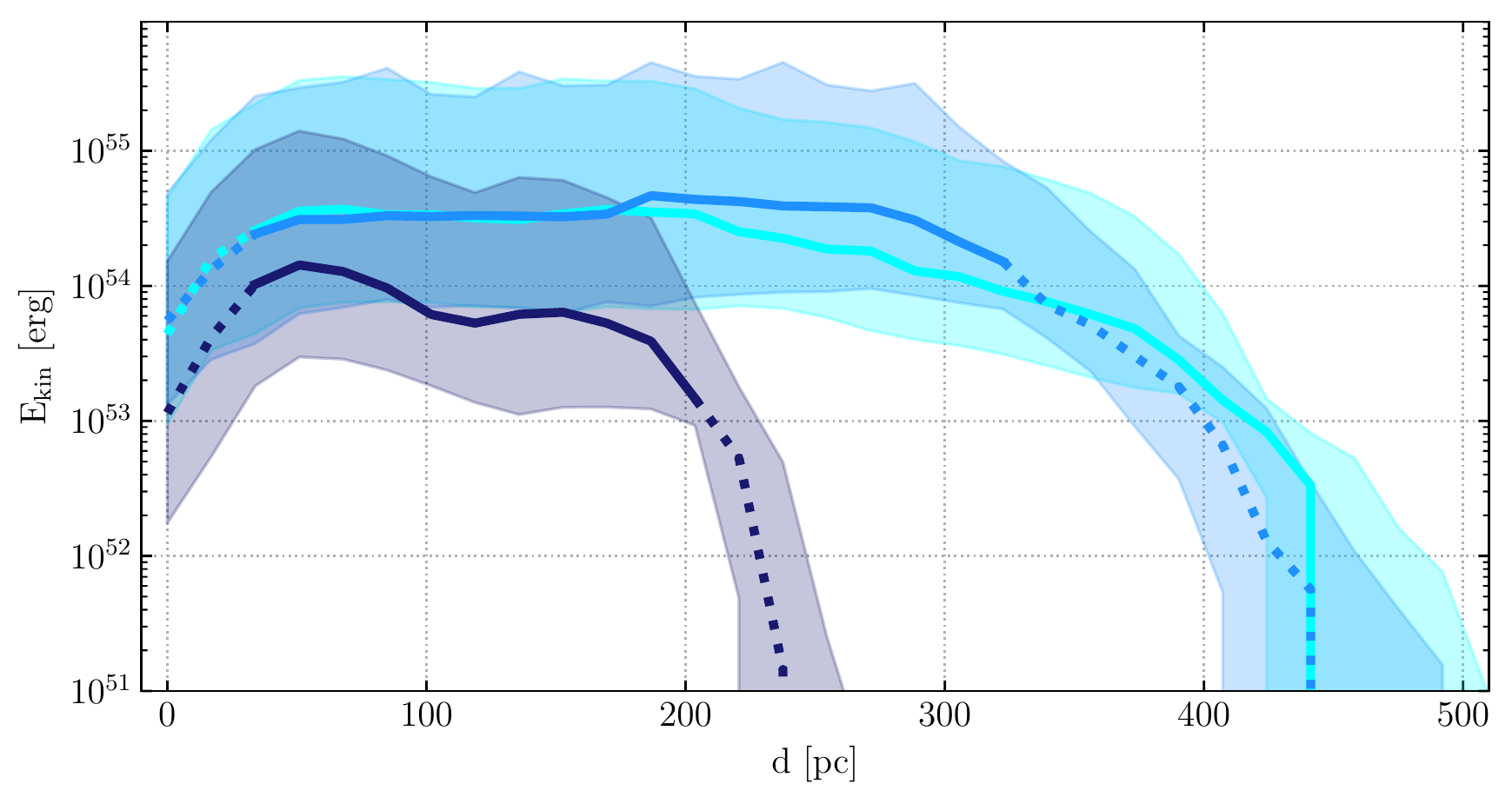}~~
    \includegraphics[width=0.48\linewidth]{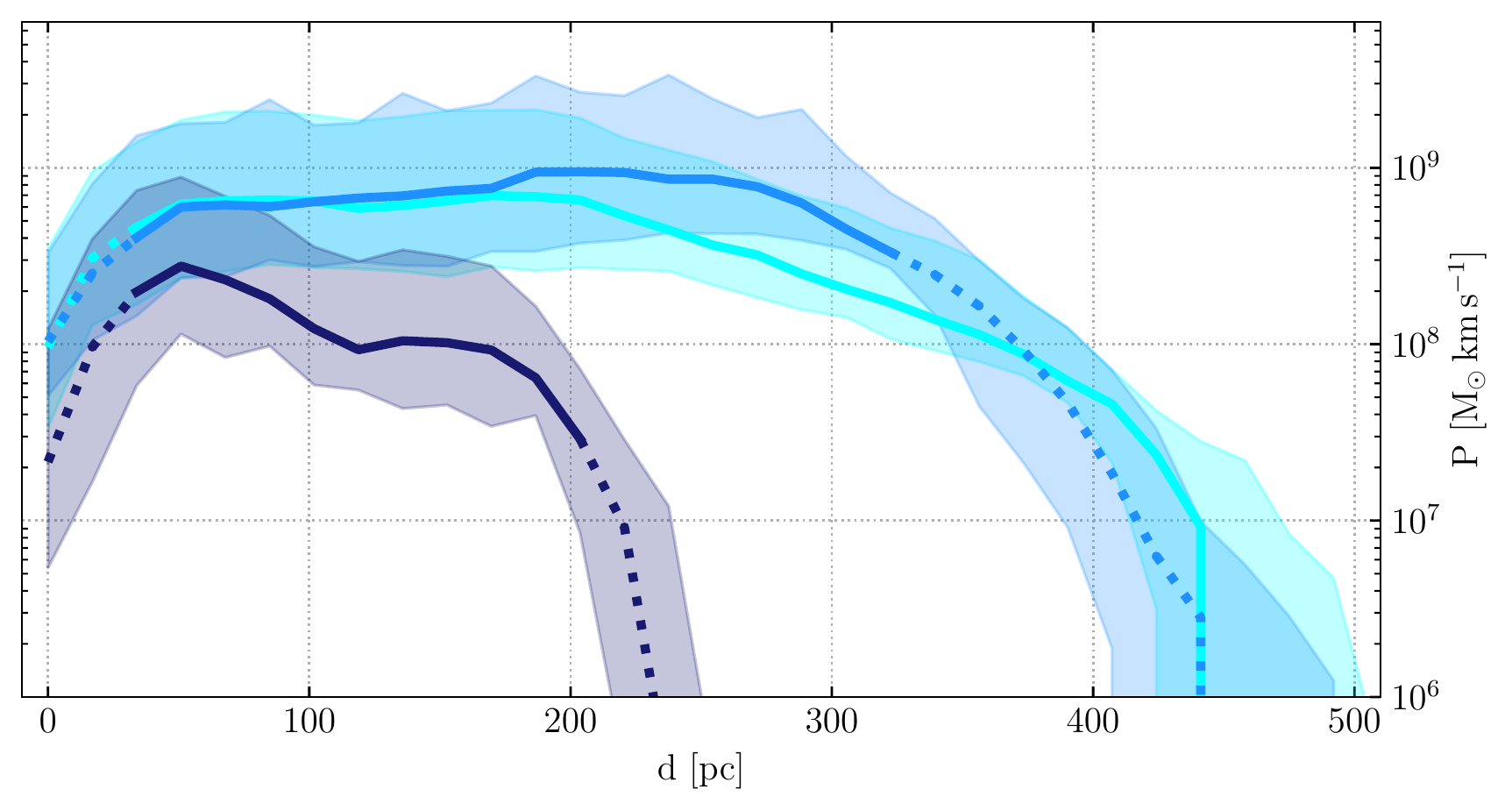}
    \caption{Deprojected kinetic energy (\emph{left}) and deprojected momentum (\emph{right}) of the molecular outflow averaged over 0.1\,Myr as a function of time since ejection (\emph{top}) and as a function of deprojected distance between outflow and launching site (\emph{bottom}). The top panel implicitly assumes continuous mass ejection without accelerations to the gas after ejection,  while the lower panel assumes approximately constant starting mass outflow rate over the lifetime of the starburst. The shaded area indicates approximate errors ($16^{th}$ to $84^{th}$ percentile), which are dominated by uncertainties in the deprojection geometry. Dotted lines represent the ranges where confusion with gas in the disk occurs and where the limited field-of-view affects the completeness.}
    \label{figure: outflow energy momentum}
\end{figure*}

Similar to mass outflow rate, energy and momentum can be calculated as a function of time and distance which is shown in figure~\ref{figure: outflow energy momentum}. In equations~\ref{equation: outflow rate time} and \ref{equation: outflow rate distance} the molecular outflow rate $\dot{m}_i$ is replaced by kinetic energy $E_{\mathrm{kin},i} = \frac{1}{2} m_i\,v_i^2$ or momentum $P_i = m_i\,v_i$. As with the outflow rate, the dominant sources of error are the uncertainty in the launching site and the geometry for which we Monte Carlo the errors as described before. Our median estimate with 16$^{th}$ and 84$^{th}$ percentile uncertainties are given below and in table~\ref{table: results}.

The kinetic energy in the outflow integrated over the past 1.0\,Myr is $3.9 \times 10^{54}$\,erg ($^{+0.91}_{-0.75}$\,dex) in \co10, $4.5 \times 10^{54}$\,erg ($^{+0.94}_{-0.80}$\,dex) in \co21, and $6.5 \times 10^{53}$\,erg ($^{+0.58}_{-0.83}$\,dex) for \co32. Within 20\arcsec\ (340\,pc), the kinetic energies amount to $2.5 \times 10^{54}$\,erg ($^{+0.96}_{-0.65}$\,dex) in \co10, $3.1 \times 10^{54}$\,erg ($^{+0.96}_{-0.65}$\,dex) in \co21, and $4.3 \times 10^{53}$\,erg ($^{+0.98}_{-0.64}$\,dex) for \co32. For the reasons described above, the \co32 measurement is a lower limit, thus the results for the lower transitions are  consistent.

NGC\,253 does not appear to host an energetically important AGN, and the outflow is driven by the starburst. It is interesting then to compare our results for the kinetic energy to the energy released by the starburst. We assume the current star formation rate of $\sim 2.8$\,\Msunyr in the central region \citep{Ott:2005il,2015MNRAS.450L..80B} that has been approximately constant over the last few Myr.

The total energy $E_\mathrm{bol}$ produced by the starburst is simply the time integrated bolometric luminosity $L_\mathrm{bol}$ which depends\footnote{We follow IAU resolution B2 that defines the bolometric magnitude in absolute terms and eliminates the dependence on the variable magnitude of the sun.} on the bolometric magnitude $M_\mathrm{bol}$.
\begin{eqnarray}
    E_\mathrm{bol} &=& L_\mathrm{bol} \times \Delta t\\
    &=& 3 \times 10^{35-0.4\,M_\mathrm{bol}} \Delta t \left( \frac{SFR}{1\,\mathrm{M}_\odot\,\mathrm{yr}^{-1}} \right)\,\mathrm{erg\,s}^{-1}
    \label{equation: bolometric energy}
\end{eqnarray}
According to Starburst99 \citep[figure~46 in][]{Leitherer:1999jt}, the bolometric magnitude of a starburst at an age of $10^7$\,yr to $10^8$\,yr is $M_\mathrm{bol} \sim -20.5$ for $\mathrm{SFR} = 1$\,\Msunyr.
The total energy output of the starburst over the past 1\,Myr is thus $4.2 \times 10^{57}$\,erg. The observed kinetic energy $\sim3.9-4.5\times10^{54}$\,erg in the outflow is a factor of $\sim 10^3$ lower which places the coupling efficiency of outflow kinetic energy to starburst energy at $\sim 0.1$\%.

In terms of only kinetic energy, the fraction is higher. Primarily supernovae and winds supply kinetic energy to the ISM which can be estimated from the energy deposition rate according to \citet{Leitherer:1999jt} as given in \citet{Chisholm:2017bu} and \citet{Murray:2005jt}.
\begin{eqnarray}
    \dot{E}_\mathrm{SN} &= 3 \times 10^{41} \left( \frac{SFR}{1\,\mathrm{M}_\odot\,\mathrm{yr}^{-1}} \right) \,\mathrm{erg\,s}^{-1}
    \label{equation: supernova energy}
\end{eqnarray}
Each SN releases approximately $10^{51}$\,erg in kinetic energy, with the progenitor releasing a similar amount of kinetic energy during its lifetime by winds \citep[e.g.][]{Leitherer:1999jt}. The approximate total kinetic energy released by SNe in the past 1\,Myr is then $\sim 5.3 \times 10^{55}$\,erg, compared to the $\sim3.9-4.5\times10^{54}$\,erg we observe in the outflow. Hence, the observed starburst is sufficient to kinetically power the measured molecular outflows with $\sim 8\%$ efficiency.

The commonly adopted 50\% relative contribution of wind feedback is a first order estimate that is subject to environmental dependence and requires careful modeling to determine precisely \citep[e.g.][]{Leitherer:1999jt}. Furthermore, it should be noted that the observed outflow energy and its error is based on a fixed mass conversion factor that may vary. The uncertainty on the energy coupling efficiency is thus substantial and it should be understood as an order of magnitude comparison.

The above calculation ignores the contribution of other energies, such as the turbulent energy within the molecular outflow and the kinetic energy of the neutral and ionized gas. \citet{2009ApJ...701.1636M} derived a kinetic energy of the ionized wind in NGC\,253 of $1.3 \times 10^{53}$\,erg or more than one order of magnitude lower than the molecular outflow kinetic energy. The molecular outflow is slower ($50-100$\,\kms on the scales we observed here) than the ionized outflow \citep[up to $\sim 400$\,\kms,][]{2009ApJ...701.1636M} but also more massive. The ionized outflow thus has only a very small effect on the total kinetic energy and the coupling efficiency.

Deprojected outflow momenta integrated over the past 1.0\,Myr are $6.9 \times 10^8$\,\Msunkms ($^{+0.50}_{-0.49}$\,dex), $8.7 \times 10^8$\,\Msunkms ($^{+0.57}_{-0.57}$\,dex) and $1.2 \times 10^8$\,\Msunkms ($^{+0.33}_{-0.59}$\,dex) for \co10, (2--1) and (3--2), respectively. Within 20\arcsec\ (340\,pc) deprojected distance from the launching site the outflow momenta integrate to $4.8 \times 10^8$\,\Msunkms ($^{+0.48}_{-0.35}$\,dex) in \co10, $6.4 \times 10^8$\,\Msunkms ($^{+0.49}_{-0.34}$\,dex) in \co21 and $8.0 \times 10^7$\,\Msunkms ($^{+0.50}_{-0.39}$\,dex) in \co32.

The momentum released initially by SNe is given in \citep{Murray:2005jt}:
\begin{eqnarray}
    \dot{P}_\mathrm{SN} &=& 2 \times 10^{33} \left( \frac{SFR}{1\,\mathrm{M}_\odot\,\mathrm{yr}^{-1}} \right) \,\mathrm{g\,cm\,s}^{-2} \\
    &=& 317 \left( \frac{SFR}{1\,\mathrm{M}_\odot\,\mathrm{yr}^{-1}} \right) \,\mathrm{M_\odot\,yr}^{-1}\,\mathrm{km\,s}^{-1}
    \label{equation: supernova momentum}
\end{eqnarray}
In 1\,Myr, a constant SFR of 2.8\,\Msunyr yields $8.9 \times 10^8$\,\Msunkms. Assuming a contribution by stellar winds of the same order \citep{Leitherer:1999jt}, the total momentum is $1.8 \times 10^9$\,\Msunkms or roughly twice the observed outflow momentum.
SNe, however, gain significant amounts\footnote{Assuming a Salpeter-like IMF ($\alpha=2.35$, mass range $0.1-100$\,\Msun, $\mathrm{Z} = 0.008$). The usual uncertainties related to the shape, upper mass cutoff and influence of binary stars apply.} of momentum by sweeping up surrounding material. From simulations, the total momentum supplied to the ISM is expected to be $2.8 \times 10^5$\,\Msunkms per SNe (\citealt{Kim:2015iy} and references therein). For a constant SFR of 2.8\,\Msunyr over 1\,Myr, this amounts to $1.0 \times 10^{10}$\,\Msunkms or $2.0 \times 10^{10}$\,\Msunkms when adopting 50\% contribution by stellar winds which is about four times the observed momentum. 
The efficiency of transferring feedback momentum to outflow momentum is thus in the range $27-49$\% considering the initially available momentum or $2.5-4$\% efficiency for total to outflow momentum transfer.

These outflow momenta are much higher than the momentum currently produced by young ($<10$\,Myr) super star clusters in the starburst. \citet{2018ApJ...869..126L} list 14 candidate clusters that together produce $1.5\times10^7$\,\Msunkms measured from gas kinematics, a factor $10-100$ lower than the observed outflow momentum. The currently forming (super-) star cluster thus could not have launched the outflow but the feedback of another population of stars is needed to explain the observed outflows. This is indicative of the time delay of SF feedback.

Energy and momentum curves in figure~\ref{figure: outflow energy momentum} differ only by a factor $v$ but follow a similar evolution. This implies that the median velocity at a given distance must be roughly constant along the outflow. As the curves as a function of distance are roughly constant within $50\,\mathrm{pc}<s<300\,\mathrm{pc}$, especially for kinetic energy, the outflow mass at a given distance must also be approximately constant along the outflow.
The decline in energy and momentum below 50\,pc is caused by a decrease in outflow mass, again because both curves follow a similar trend. This is at least partially related to the difficulty of separating outflow from disk where the former emerges from the latter. The decrease could also be interpreted physically as the outflow sweeping up mass while emerging from the disk. An estimation of the relative importance of these effects requires high-resolution modelling of the outflow that are not possible yet because we do not know the detailed outflow geometry.
The drop beyond $\sim 300$\,pc ($\sim 200$\,pc in \co32) is partially related to reaching the edge of the field-of-view. Discerning this effect from an actual decrease is not possible with our data as we do not know the inclination at every location in the outflow. The edge of the field-of-view thus corresponds to a range of deprojected distances from the disk which gradually depresses the curve rather than showing a sudden drop. A physical reason for the decrease could be the destruction of the molecular gas, e.g. photo-dissociation by the intense starburst radiation or ionization.

The kinetic energy and momentum evolution in figure~\ref{figure: outflow energy momentum} thus suggest both energy and momentum conservation along the outflow from $\sim50$\,pc to $\sim 300$\,pc, as well as approximately constant molecular gas mass.

When additionally assuming no acceleration of the outflow after launch, it becomes possible to study the time evolution. The corresponding plots (figure~\ref{figure: outflow rate} top and \ref{figure: outflow energy momentum} top) all show a peak within the past 0.5\,Myr and steady decrease towards earlier gas ejection times. Corresponding to the decline towards zero distance, the decrease towards zero ejection time is most likely a methodological complication. From the peak at t$_\mathrm{eject} = 0.2-0.3$\,Myr, kinetic energy and momentum in the outflow drops by a factor of 10 within $\sim 2$\,Myr. This decline would be physically plausible if the starburst in NGC\,253 is very young and taking into account a time delay between start of star formation, feedback and efficient outflow driving (superbubble breakout). For the observed age of the starburst of $20-30$\,Myr \citep{Rieke:1980hh,Engelbracht:1998cj} this scenario is implausible. Time delays of $>20$\,Myr are longer than the lifetime of the most massive stars. A younger generation of massive stars at an age of $\sim 6$\,Myr \citep{Kornei:2009ee} may, however, drive the currently visible molecular outflows. If this were to be true, a time delay between star formation and outflow launching of $\sim 4$\,Myr is implied. Outflow launching in this context means the time after which the outflow reaches a mass loading $\eta>1$.
The time delay is 2\,Myr until the outflow carries more energy (momentum) than the feedback kinetic energy (momentum) of a single high mass star. Note that these rough estimates depend on the assumption of no acceleration (positive, nor negative) of the outflow after being launched from the disk which might be a close enough approximation on these scales of a few hundred parsecs.
The very young ($\lesssim 1$\,Myr) and still deeply embedded super star cluster discussed recently by \citet{2017ApJ...849...81A} and \citet{2018ApJ...869..126L} are most likely too young to have affected the observed molecular outflow.


\section{Summary and Conclusions}\label{section: summary}

We present \co32 observations taken with ALMA that offer an unprecedented resolution of $\sim 0.15\arcsec$ ($\sim 2$\,pc) in the starbursting center of NGC\,253.
The new high resolution data show structures consistent with previous lower resolution observations in other CO lines,  revealing the complexity of the molecular ISM in a starburst on scales of a few parsecs. 

We use archival \co10, \co21, and the new \co32 ALMA observations to perform a position-position-velocity decomposition of the emission into different structures. The bulk of the emission is associated with a rotating disk with streaming motions due to the bar. The rest of the emission is incompatible with a simple kinematic model of a disk plus a bar. This ``non-disk'' component is further decomposed into an outflow, an expanding superbubble (part of which may be associated with outflowing gas) and a potential second kinematic component within the disk.

We find CO line luminosities of the disk component of $2.8\times10^8$\,\Kkmspc, $2.3\times10^8$\,\Kkmspc and $1.8\times10^8$\,\Kkmspc for \co10, (2--1) and (3--2), respectively. The fractional luminosity of the non-disk component is small, amounting the $\sim7-16\%$ of the total.  A significant amount of the outflow emission we identify is faint and diffuse, while part of the emission is in discrete, higher surface brightness structures (e.g., the SW streamer).

Assuming a starburst conversion factor, we estimate the molecular gas mass from the three CO transitions. Masses match within 10\% for the disk component and within 50\% for the non-disk component. The total gas mass in the center of NGC\,253 is $\sim 3.6 \times 10^8$\,M$_\odot$, with $\sim 0.5 \times 10^8$\,M$_\odot$ in the non-disk component.

We further estimate the deprojected molecular mass outflow rate, kinetic energy and momentum in the starburst of NGC\,253. The observed gas distribution can be interpreted to have formed in two ways: (1) by constant starting mass outflow rate over the lifetime of the starburst and (2) through continuous gas ejection without acceleration of the gas after ejection. In the first interpretation, the molecular mass outflow rate averaged over a deprojected distance of 340\,pc (20\arcsec) from the launching site is $29-39$\,\Msunyr. Typical uncertainties are 0.4\,dex. The majority of this outflow rate is contributed by massive localized features such as the SW/SE streamers, with a significant contribution by diffuse molecular gas. The mass loading factor $\eta=\dot{M}_\mathrm{SFR}/\dot{M}_\mathrm{out}\sim14-20$ is relatively high. Due to the limited field-of-view of our observations, this $\eta$ applies to gas ejected as far away as 340\,pc: the fraction of mass that makes it to the far regions of the halo or escapes is not known. 
The kinetic energy of the molecular outflow within 340\,pc from the launching site is $2.5-3.1 \times 10^{54}$\,erg with a $\sim 0.8$\,dex error. The coupling efficiency of kinetic energy in the outflow to the total energy released by the starburst is $\sim 0.1$\% while the coupling to only the kinetic energy is higher at $\sim 8$\%. Including other phases of the outflow would increase this efficiency. The kinetic energy of the ionized outflow is negligible relative to the molecular outflow.
The outflow momenta within the same distance are $4.8-6.4 \times 10^8$\,\Msunkms (error $\sim 0.5$\,dex) which is $\sim 2.5-4$\% of the momentum supplied by SNe and winds. 
These best estimates for the physical properties of the outflow are derived from the \co10 and (2--1) observations. The very high resolution of the \co32 data is necessary to identify the outflow features that connect to the central regions.

When interpreting the outflow as a structure of constant velocity along the outflow, the time evolution can be reconstructed. We derive outflow rate, kinetic energy and momentum within the approximate dynamical time scale of 1\,Myr and find lower values compared to the previous interpretation. The difference is systematic at the $\sim 30-40$\% level. The outflow rate is $14-20$\,\Msunyr (0.3\,dex), kinetic energy $2.5-3.1 \times 10^{54}$\,erg (0.8\,dex) and momentum $4.8-6.4 \times 10^8$\,\Msunkms (0.5\,dex). 

For all measurements given above, we assume a fixed starburst mass conversion factor of $\mathrm{X}_{\mathrm{CO}} = 0.5\times10^{20}\,\left(\mathrm{K\,km\,s}^{-1}\right)^{-1}$. The quoted uncertainties are primarily systematic due to the unknown geometry of the outflow and its launching sites. A further uncertainty of $30-40$\% ($\sim 0.1$\,dex) comes from the assumptions regarding the outflowing material (constant starting mass over the lifetime of the starburst  vs.\  continuous gas ejection without acceleration). These limitations need to be addressed in the future. In principle, ALMA can provide the very high resolution and sensitivity needed to enable this detailed view of a starburst also on larger scales than probed in this study.


\software{CASA \citep{McMullin:2007tj}, astropy \citep{Collaboration:2013cd,Collaboration:2018ji}, APLpy \citep{Robitaille:2012wl}}

\acknowledgements
We would like to thank the anonymous referee for their constructive feedback which helped improve the paper.
The authors thank Jan-Torge Schindler and Roberto Decarli for insightful discussion and advise.
This paper makes use of the following ALMA data: ADS/JAO.ALMA \#2011.1.00172.S, \#2012.1.00108.S, and \#2015.1.00274.S. ALMA is a partnership of ESO (representing its member states), NSF (USA) and NINS (Japan), together with NRC (Canada), NSC and ASIAA (Taiwan), and KASI (Republic of Korea), in cooperation with the Republic of Chile. The Joint ALMA Observatory is operated by ESO, AUI/NRAO and NAOJ.
The National Radio Astronomy Observatory is a facility of the National Science Foundation operated under cooperative agreement by Associated Universities, Inc.
Part of the work presented in this paper was carried out at the Finnish Center for Astronomy with ESO (FINCA).
The work of AKL is partially supported by NASA ADAP grants NNX16AF48G and NNX17AF39G and National Science Foundation under grants No.~1615105, 1615109, and 1653300.


\FloatBarrier
\newpage
\appendix

\section{Kinematic model of the central molecular gas}\label{appendix: model}

We derive a model for the velocity of the disk component from the \co10 observations using the kinematic fitting tool \texttt{diskfit} \citep{2007ApJ...664..204S,2010MNRAS.404.1733S,2015arXiv150907120S}. As mentioned in section~\ref{subsection: ppV separation}, these models benefit from large area which is why we base them on the \co10 observations. A $20\sigma$ threshold ensures that the model is fitted to the bright disk excluding any fainter outflows.

In \texttt{diskfit}, we fit using the velocity field fitter. The names of the set options in \texttt{diskfit} are given in paratheses in the following. The model is fitted to all pixels within an ellipse of $75\arcsec$ major axis length (\texttt{regrad}), $\mathrm{PA} = 53^\circ$ (\texttt{regpa}) and ellipticity $\epsilon = 0.66$ (\texttt{regeps}). Outside this range, we use a sampling factor of 2 pixels (\texttt{istepout}). During the fit, the center is held fixed while we fit for disk position angle and ellipticity with initial guesses of $\mathrm{PA} = 53^\circ$ and $\epsilon = 0.66$ based on by eye inspection (line 9 and 10 of the parameter file). We allow the model to fit for non-axisymmetric flows with $\mathrm{PA} = 78^\circ$ initial guess and order $m=2$ (line 12) which means \texttt{diskfit} will fit for rotation plus a bisymmetric model with $m=2$ perturbations to the potential (bar). As the \co10 data cover the kinematic center we set the inner interpolation toggle to true which assumes the velocity raises linearly within the innermost fitted ring. We do not fit for radial flows (radial flows toggle) because it allows to many degrees of freedom and produces bad models as is warned about in the \texttt{diskfit} manual. We further fit for the systemic velocity and exclude warps from the model. A model with these parameters is fitted in rings at radii $12.5\arcsec$, $25\arcsec$, $37.5\arcsec$, $50\arcsec$, $62.5\arcsec$, $75\arcsec$, $87.5\arcsec$, $100\arcsec$, $125\arcsec$, $150\arcsec$, $175\arcsec$, $200\arcsec$, $225\arcsec$, $250\arcsec$, $275\arcsec$ and $300\arcsec$.

The residuals show a slight mismatch in velocity of $\sim 20-30$\,\kms along the direction of the bar. This is likely due to the bar being underestimated because the \co10 image covers only the inner half of the total extent of the bar. The mismatch gets larger when fitting a model to the smaller images of \co21 and (3--2) which confirms that it is caused by lack of observed area. We fit this mismatch in the velocity field with an additional 2D Gaussian component and add it to the velocity field of the diskfit model to obtain a better model. Note that this additional component is not physically motivated or meaningful but purely aims to counteract the effect of limited observation area.

\begin{figure*}[h!]
	\centering
	\includegraphics[scale=0.23]{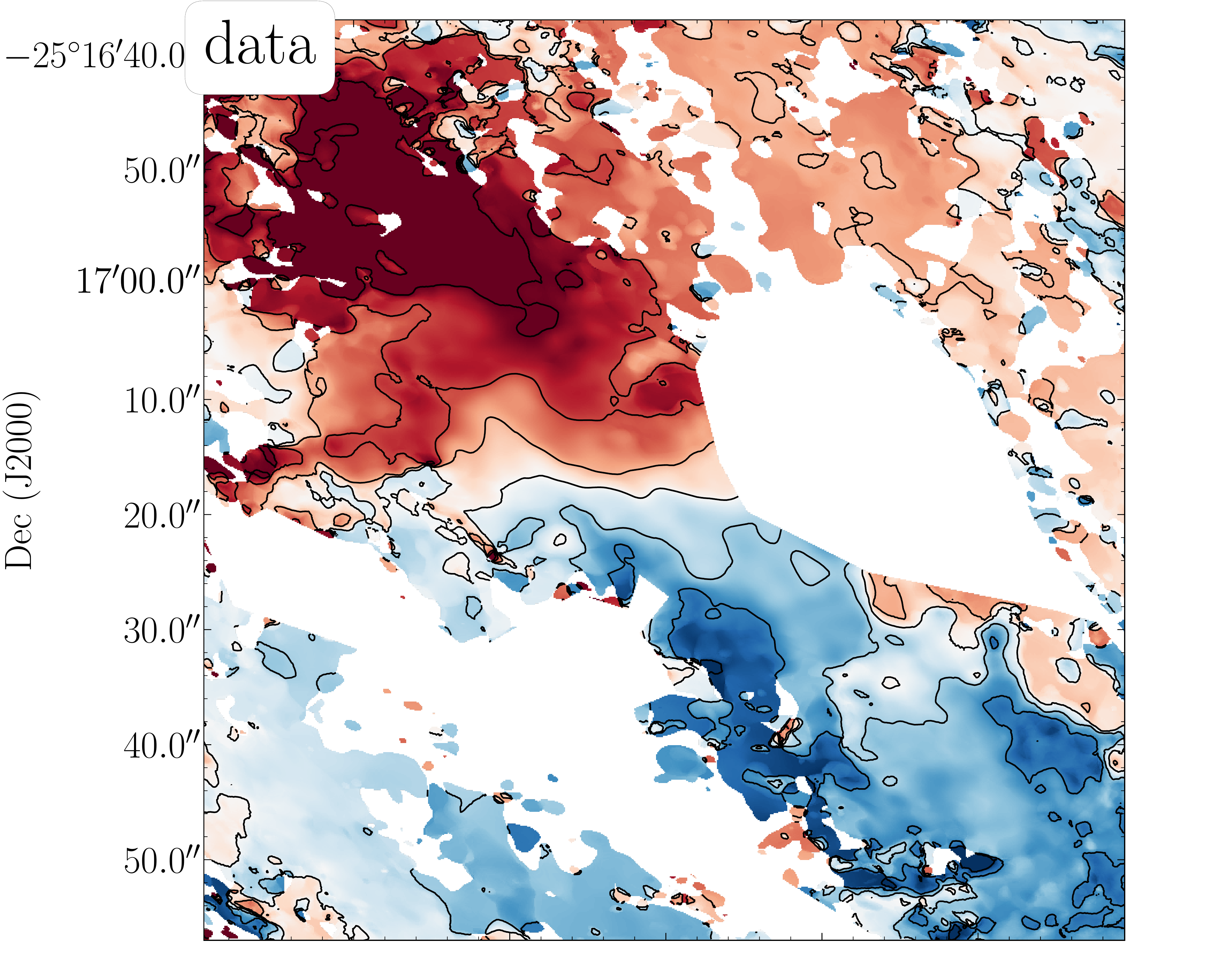}
	\includegraphics[scale=0.23]{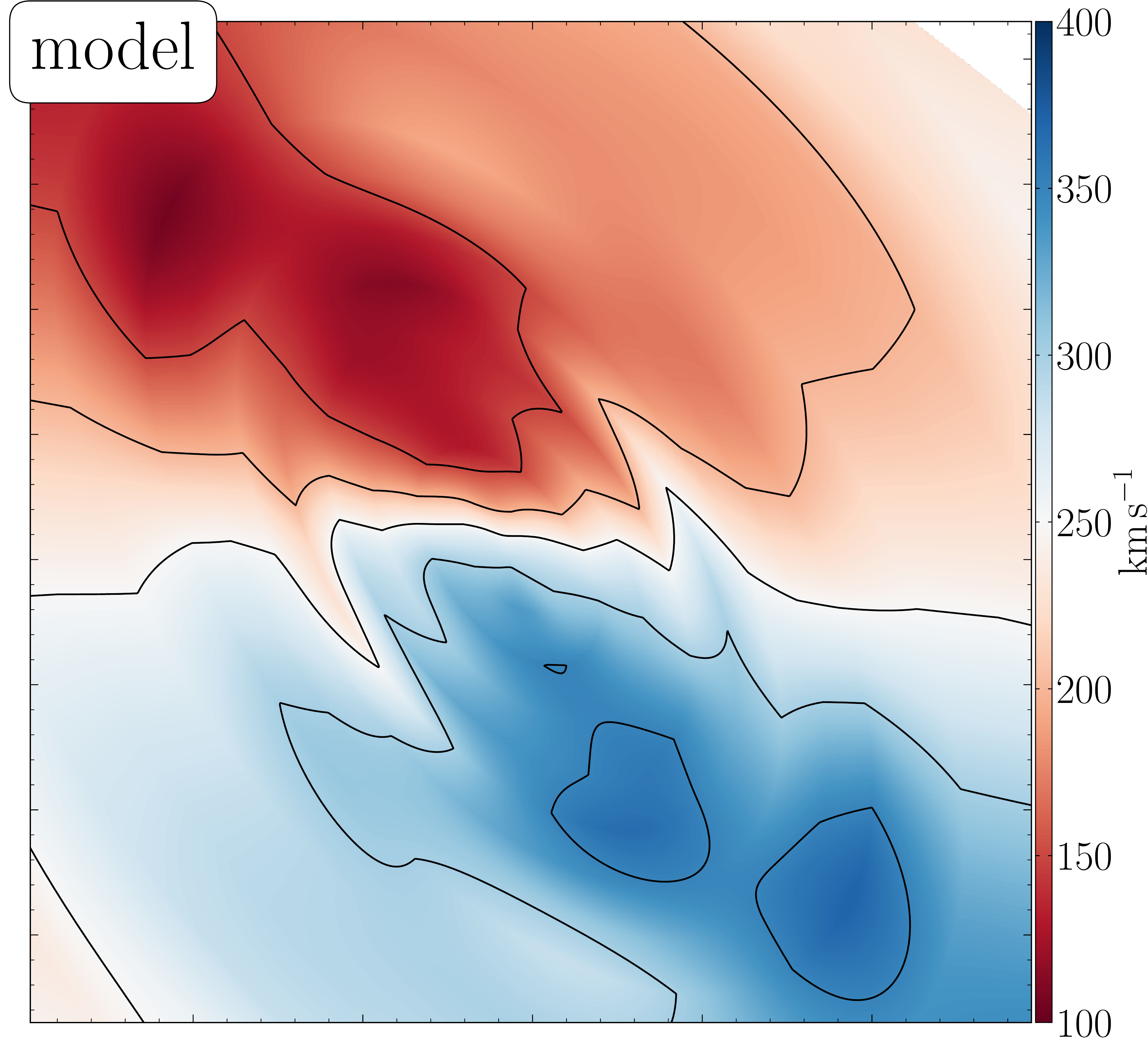}\\
	\includegraphics[scale=0.23]{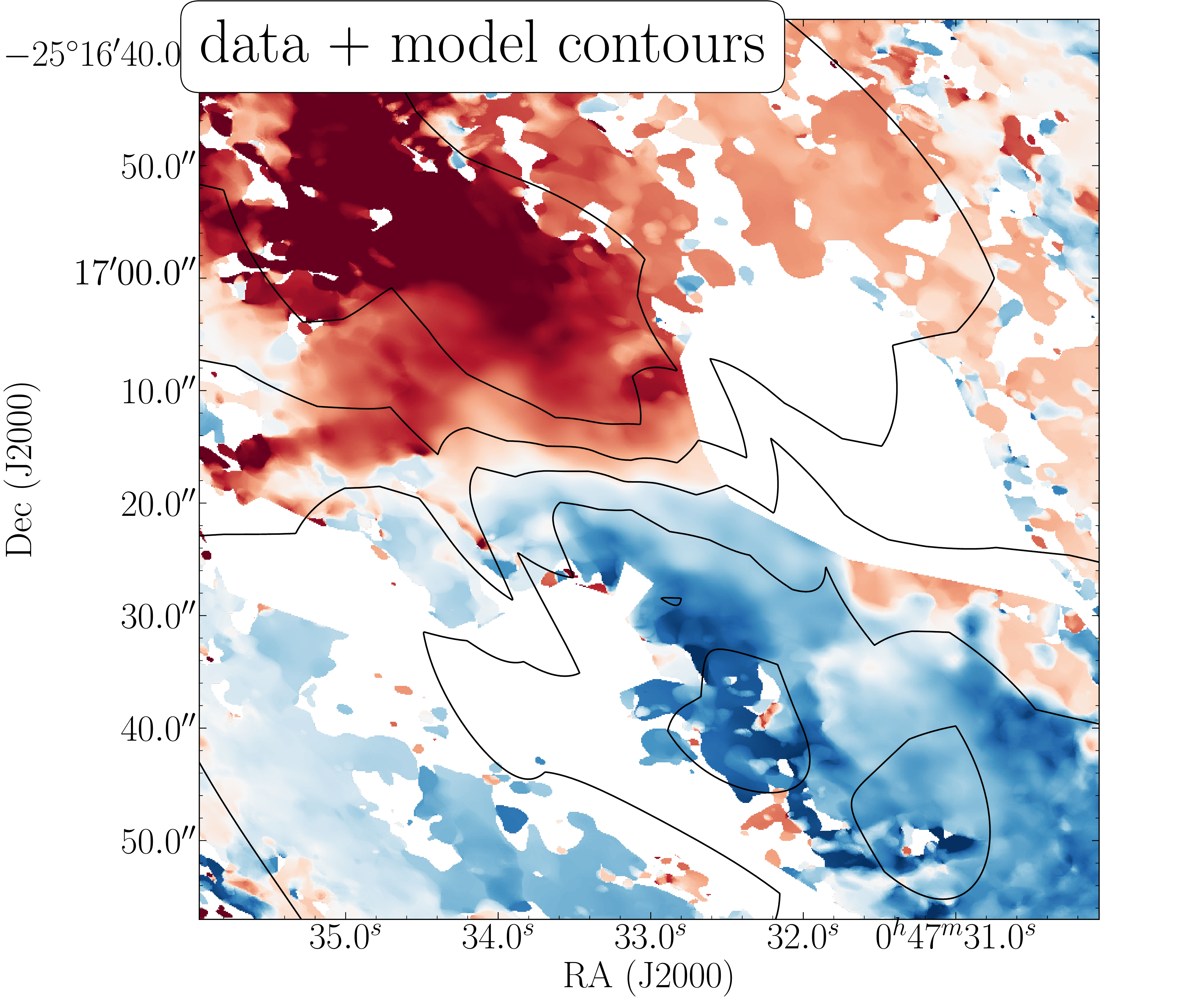}
	\includegraphics[scale=0.23]{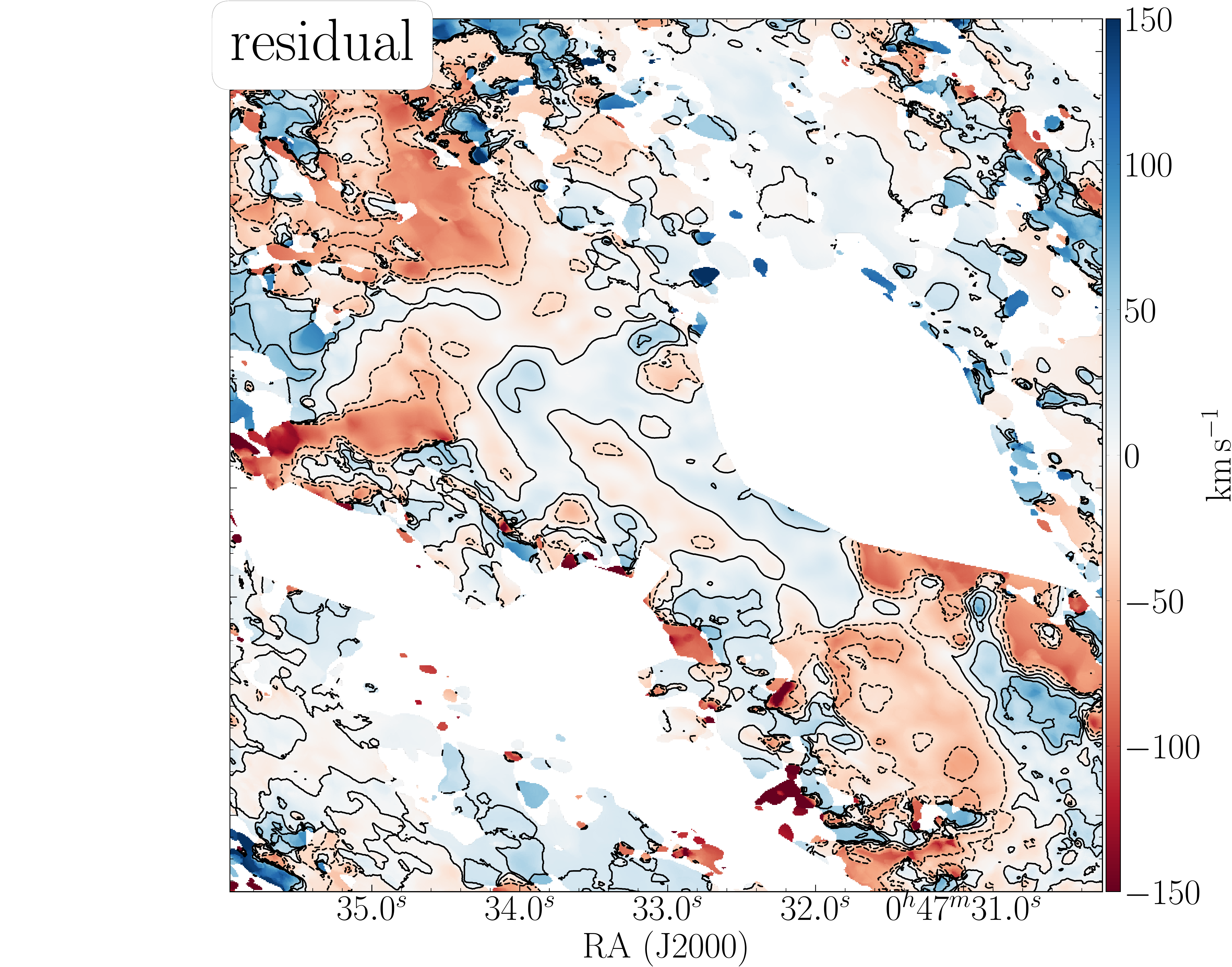}
	\caption{Input data and model represented as velocity fields. \emph{top left}: \co10 velocity field to which the model is fitted, known foreground emission is removed; \emph{top right}: model velocity field; \emph{bottom left}: \co10 velocity overlaid with model contours; \emph{bottom right}: residual velocity. The velocity fields use the same colorscale from 100\,\kms to 400\,\kms with contours in steps of 50\,\kms within that range. The residual velocity uses the same colorscale relative to the systemic velocity of 250\,\kms with contours from -50\,\kms to +50\,\kms in steps of 25\,\kms.}
	\label{figure: model}
\end{figure*}

Figure~\ref{figure: model} shows the velocity field of the model in comparison to the input \co10 velocity field. The model typically fits the observed velocity field better than $\pm 25$\,\kms; larger deviations occur mostly over small areas of order one beam size. The model thus successfully reproduces the large scale velocity field.

\FloatBarrier
\newpage
\section{Velocity width of the disk mask}\label{appendix: disk velocity range}

The definition of the disk mask is crucial for this analysis as it determines if a molecular cloud is considered kinematically consistent with the disk or if it is potentially outflowing. The position of the disk mask in ppV space is set by the disk model but the width (velocity range $\Delta v$) of the mask is a free parameter. From figure~\ref{figure: all pV diagrams} it is obvious that $\Delta v$ depends on the distance from the major axis which is most simply accounted for by a parametrisation of form $\Delta v = a \exp \left( -\left( \frac{x}{b} \right)^2 \right) +c$. Finding the best fit values for $a, b$ and $c$ is difficult to do mathematically as the fit would need to be on the disk component that we want to determine from the mask first. We therefore select values that visually fit the pV diagrams as best as possible. They are given in equation~\ref{equation: delta v}.

\begin{figure}[h!]
    \centering
    \includegraphics[width=\textwidth]{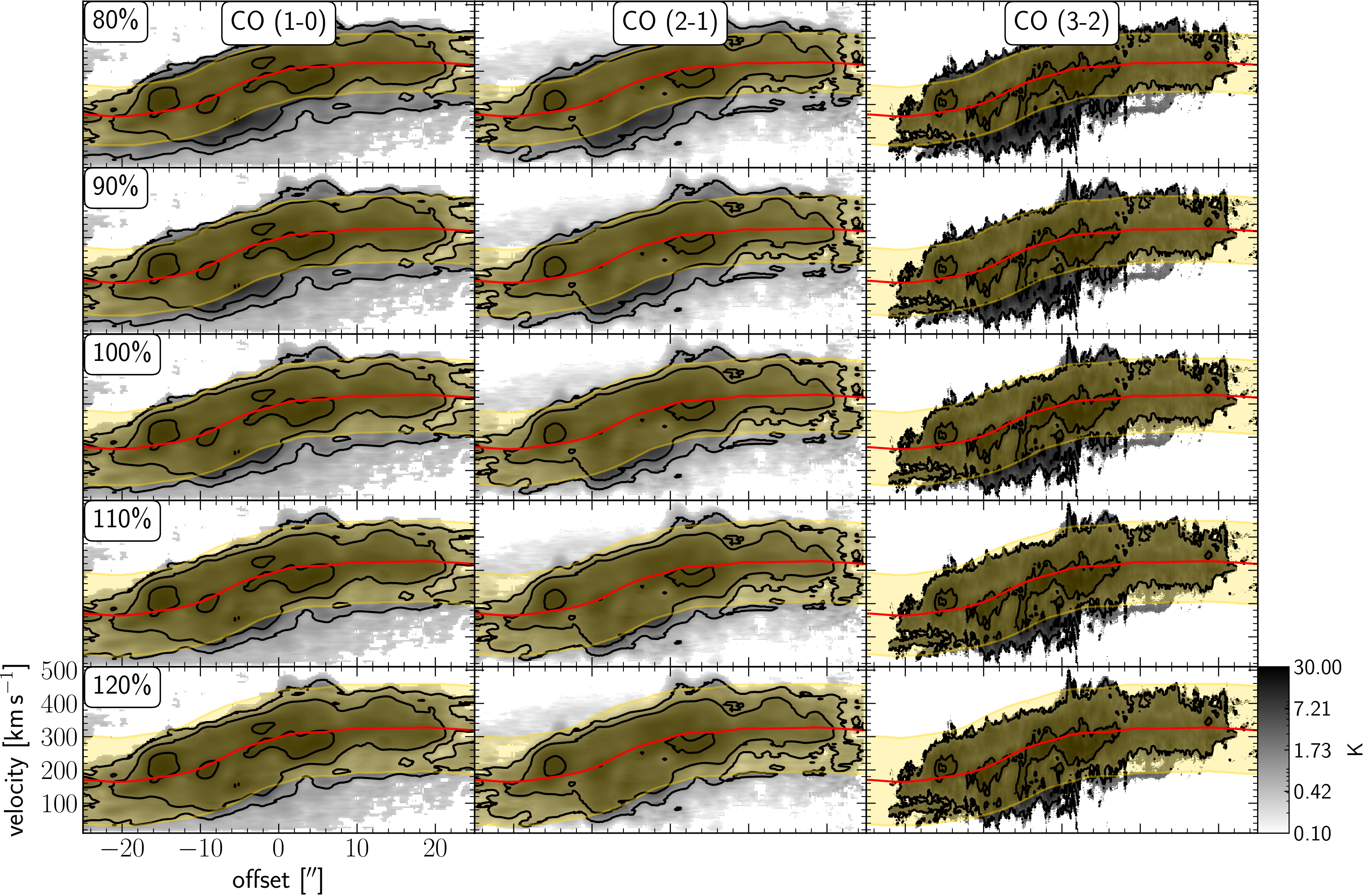}
    \caption{Position-velocity diagrams for the central slice along the major axis (offset 0.0\arcsec) overlaid with five different choices for the velocity width of the disk mask. Greyscale and overlays are identical to figure~\ref{figure: sample slice} and \ref{figure: all pV diagrams}. The central panel shows the visually best fitting mask as defined by equation~\ref{equation: delta v}. The other panels use masks of 80\%, 90\%, 110\% and 120\% of the best fit width. Even 10\% change in mask width lead to noticeable mismatch in the high-resolution \co32 data.}
    \label{figure: disk mask width}
\end{figure}

Figure~\ref{figure: disk mask width} shows five alternative masks that vary only in width by 10\% from the best fit mask. It is apparent that even slight changes of 10\% deteriorate the fit between mask and disk emission. More narrow masks obviously do not cover all disk emission whereas the wider masks include spike features that are kinematically inconsistent with disk rotation.

{}

The effect of a $^{+10}_{-10}$\% ($^{+0.04}_{-0.04}$\,dex) variation in mask velocity width results in a $^{+0.01}_{-0.01}$\,dex change in integrated disk luminosity for \co10, \co21 and \co32. The non-disk components is less bright than the disk and thus shows a higher relative variation when changing the mask width: luminosities vary by $^{-0.07}_{+0.08}$\,dex ($^{-0.06}_{+0.07}$\,dex, $^{-0.11}_{+0.12}$\,dex) for \co10 (\co21, \co32). Note the inverse scaling between disk and non-disk due to shifting the balance between the two components for a constant total luminosity. To first order, the same percentage changes apply to the further quantities mass, outflow rate, energy and momentum.


\FloatBarrier
\newpage
\section{Disk/non-disk separation: Position-velocity diagrams}\label{appendix: all pVs}

Figure~\ref{figure: all pV diagrams} shows all pV diagrams for the slices defined in figure~\ref{figure: slice positions}. For the discussion of these diagrams, see section~\ref{section: disk separation}.

\begin{figure*}[h!]
	\centering
	\includegraphics[height=0.85\textheight]{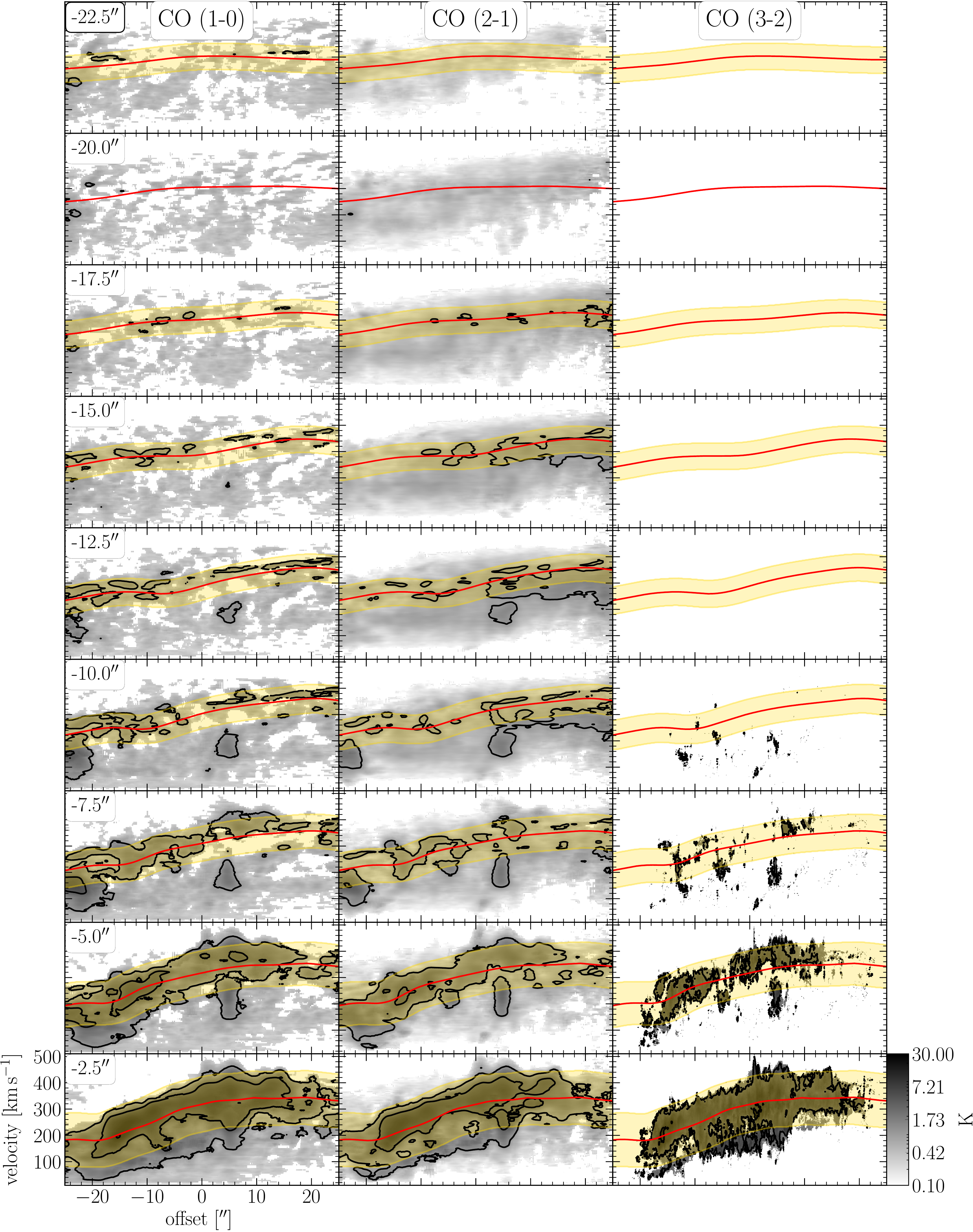}
	\caption{Position-velocity diagrams for all slices defined in section~\ref{section: disk separation} and figure~\ref{figure: slice positions} for \co10, \co21 and \co32 in the left to right column. A description of the overlays is given in Fig.~\ref{figure: sample slice}.}
	\label{figure: all pV diagrams}
\end{figure*}

\begin{figure*}[h!]
	\ContinuedFloat
	\centering
	\includegraphics[height=0.95\textheight]{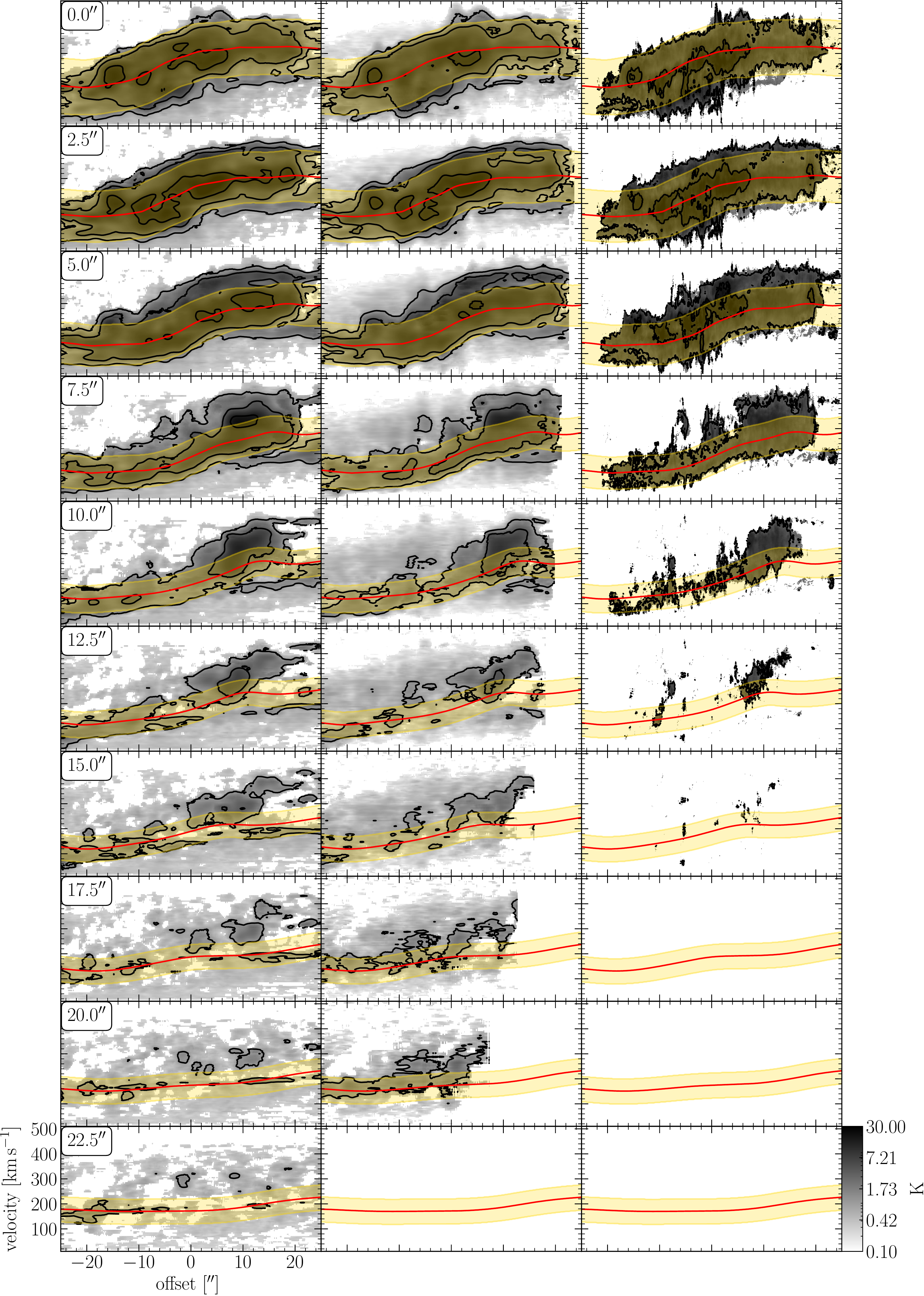}
	\caption{continued}
\end{figure*}


\FloatBarrier
\newpage
\bibliography{bibliography.bib}

\end{document}